\documentclass[preprint]{aastex631}

\begin{document}

\title{On the Spatial Distribution of Luminous Blue Variables, B[e] Supergiants, and Wolf-Rayet stars in the Large Magellanic Cloud}

\correspondingauthor{John C. Martin}
\email{jmart5@uis.edu}

\author[0000-0002-0245-508X]{John C. Martin}
\affiliation{University of Illinois Springfield}

\author[0000-0003-1720-9807]{Roberta M. Humphreys}
\affiliation{Minnesota Institute for Astrophysics, University of Minnesota}

\author[0000-0003-0221-788X]{Kris Davidson}
\affiliation{Minnesota Institute for Astrophysics, University of Minnesota}



\begin{abstract}
	We examine the spatial distributions of LBVs, B[e] supergiants, and W-R stars in the LMC, to clarify their relative ages, evolutionary states, and relationships.  This survey employs a reference catalog that was not available for previous work, comprising more than 3900 of the LMC's most luminous stars.   Our analysis shows that LBVs,  B[e] supergiants, and WR's have spatial distributions like normal stars with the same spectral types and luminosities.  Most LBVs are not isolated, nor do they require binary or multiple status to explain their spatial relationship to other populations. There are two likely exceptions:  one lower-luminosity LBV and one LBV candidate are relatively isolated and may have velocities that require additional acceleration.  The B[e] supergiants are spatially and kinematically more dispersed than LBVs, suggesting that they belong to an older population. The most luminous early-type WN's are most closely associated with the  evolved late O-type supergiants. The high luminosity late WNs, and WNh stars however, are highly concentrated in the 30 Dor region which biases the analysis.  The less luminous WNs and WCs are associated with a mix of evolved late B, A-type, and yellow supergiants which may be in a post-red-supergiant phase.  Spatial distributions of the less luminous WN, WC, and WN3/O3 stars reinforce proposed evolutionary links among those subtypes.  Our analysis also demonstrates the importance of using a comprehensive census, with reference populations clearly defined by spectral type and luminosity, and how small number statistics, especially combined with spatial clustering, can invalidate some commonly-cited statistical tests.
\end{abstract}

%% https://urldefense.com/v3/__https://astrothesaurus.org__;!!DZ3fjg!5Dmprq-SuVAGZc4bKpQ_2hx3-6WsnfAobij-hllxky2La9g13EBHOHbjRkm8QE0HhlPsHzOBiUgRzfQ$ 


\section{Introduction} \label{sec:intro}

The important final stages in the evolution of massive stars, whether they end as supernovae or collapse directly to black holes, remain uncertain. The majority of massive stars (9 -- 40 M$_{\odot}$) will pass through the red supergiant stage. For most, this is their terminal phase. However, those with initial mass greater than about 40 M$_{\odot}$, above the empirical upper luminosity boundary, do not evolve to the red supergiant stage \citep{1979ApJ...232..409H,1994PASP..106.1025H}.  Many terminal explosions exhibit interaction with circumstellar material from prior mass loss, even high mass loss events. But understanding the association of the probable SNe progenitors with the classes or groups of luminous evolved stars known for high mass loss is complicated by the latter's relative scarcity and their different evolutionary histories.  They include the Luminous Blue Variables (LBVs), B[e] supergiants (B[e]sg), Wolf-Rayet stars (WR), the yellow and red hypergiants, and the rare giant eruptions.

\vspace{2mm}

Various authors have argued about the evolutionary state of LBVs or S Dor variables and their relationship to other evolved massive stars, including the origin of their underlying instability. \citet{2015MNRAS.447..598S} suggested that LBVs are physically isolated from the population of young O-type stars in the Magellanic Clouds, and concluded that they are  runaway stars that have gained mass in binary systems and have been accelerated. \citet{2016ApJ...825...64H} and \citet{2016arXiv160802007D} tested their assumptions and showed that the confirmed LBVs are associated with  evolved massive stars. The Smith and Tombleson sample was a mixed population. When the non-LBVs were removed, the classical LBVs were associated with later O-type stars, while the less luminous LBVs were associated with more evolved supergiants. \citet{2018AJ....156..294A} recognized the incompleteness of the O and B star reference populations in the LMC and used a photometrically selected sample of bright blue stars. They similarly concluded that LBVs are not more dispersed from massive stars but are spatially distributed like the WR stars. 

\vspace{2mm}

These studies affirm the need for a more comprehensive reference sample of spectroscopically selected luminous stars for comparison, which can be separated by spectral type and luminosity. In our own galaxy, dust and gas in the Galactic plane obscure most of the massive star population from view and even with the advances from Gaia, distances to stars in the Milky Way may be uncertain.  In Local Group galaxies like the LMC, SMC, M31 and M33, the stars are essentially at the same distance. These star forming galaxies with large numbers of young, massive stars, present a more complete picture of the population of the upper HR Diagram. 

\vspace{2mm} 

Fortunately, the LMC is well-studied in this respect with many spectroscopic and photometric surveys which when combined yield a formidable sample for analyzing the spatial distribution of evolved luminous stars.  Previous studies mostly excluded the 30 Doradus region due to extreme crowding and uncertain extinction.   In \citet{2023AJ....166..214M} we presented a spectroscopically-selected HR-Diagram for the most luminous stars in the LMC combining data from many surveys including the VFTS survey of 30 Doradus \citep{2011AA...530A.108E} plus the evolved A-type supergiants, and the yellow (YSGs) and red supergiants (RSGs). 

\vspace{2mm}

\begin{figure}[t!]
\epsscale{0.9}
\plotone{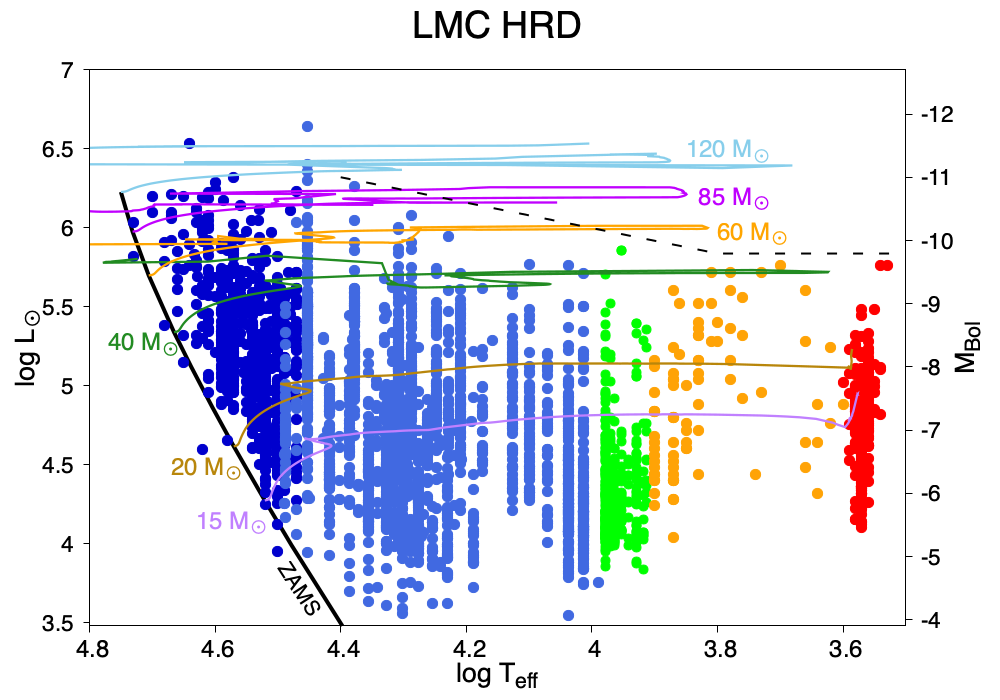}
\caption{The Hertzsprung-Russell diagram for the most luminous and massive stars in the Large Magellenic Cloud.  The Zero Age Main Sequence (ZAMS) is noted with a solid black line.  The Humphreys/Davidson limit \citep{1979ApJ...232..409H,1994PASP..106.1025H} is noted with the dashed black line.  Stars are color coded:  O-type = dark blue, B-type = light blue, A-type = green, YSG = orange, RSG = red.  The vertical striping is due to the spectroscopic temperature scale we have adopted \citep{2023AJ....166..214M}. ZAMS and evolutionary tracks are adopted from \citet{2021A&A...652A.137E} Z= 0.006 non-rotating models which is consistent with the range of spectroscopic abundances measured for luminous stars in the LMC \citep{2022A&A...658A..29R,2019PASP..131b4101A,2008AJ....136..375M}.
\label{fig:hrd}}
\end{figure}

Several additional early O-type and later type stars from the VFTS survey have since  been added to  the lists in our 2023 paper. They are included in Appendix \ref{hrdupdate}. The HR-Diagram with these new stars is shown in Figure \ref{fig:hrd}.

\vspace{2mm}

We use this catalog of 3910 luminous stars of all spectral types to investigate the spatial distributions of the LBVs, B[e]sg and WR stars.   In the following sections, we describe our selection of the samples, their spatial distributions, and our nearest-neighbor analysis. We summarize our results based on their spatial distributions and their kinematics in \S {4} and highlight our conclusions in the last section.

\vspace{2mm}

\section{Luminous Blue Variables, B[e] Supergiants, and Wolf-Rayet Stars in the LMC}

\subsection{LBVs}\label{sec:lbv}

LBVs are a relatively small group of evolved massive stars recognized by their characteristic S Dor spectroscopic and photometric variability. In its enhanced mass loss phase, i.e the LBV eruption or maximum-light phase, the star increases in visual brightness by 1 to 2 magnitudes. Its spectrum changes from its hot, quiescent state to a cooler type resembling a late A to F-type supergiant ($T_{eff} \sim$ 7000 -- 8500 K)  due to its expanded photosphere or optically thick wind. See the HR-Diagrams in \citet{1994PASP..106.1025H} and \citet{2016ApJ...825...64H} for examples of their apparent transits at nearly constant luminosity. The star may stay in this state of enhanced mass loss for several years. During their quiescent state, LBVs lie along what is known as the LBV or S Dor instability strip \citep{1981AA....99..351W,1994PASP..106.1025H}.  

\vspace{2mm}

It is important to avoid confusing LBV/S Dor variability with giant eruptions which significantly increase their bolometric luminosities, with higher mass loss, and shorter duration at maximum light. The latter are sometimes referred to as supernova imposters. See the recent review by \citet{2020Galax...8...10D} for a more complete discussion.

\vspace{2mm}

\begin{figure}
\epsscale{0.8}
\plotone{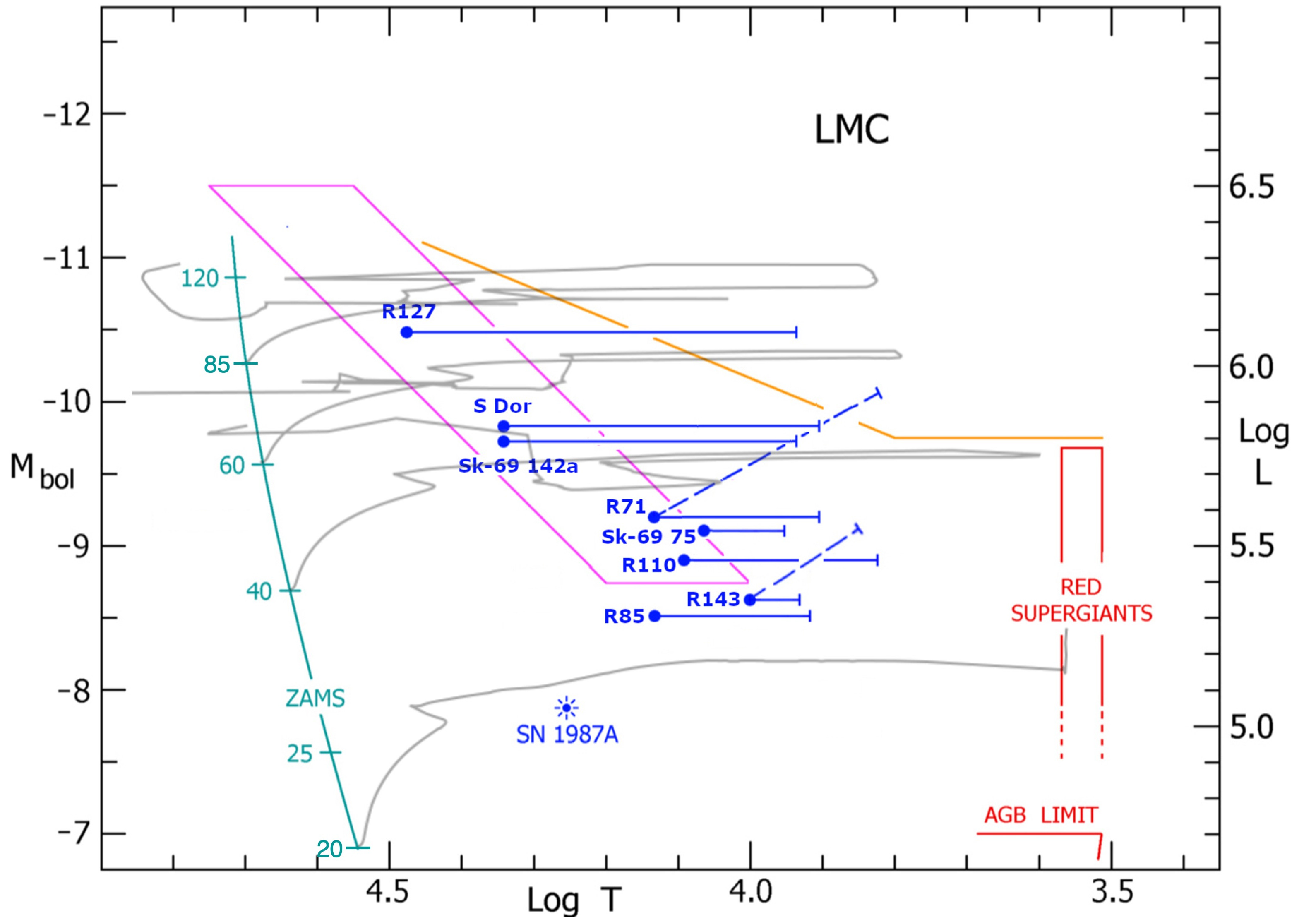}
\caption{A schematic HR diagram of the LBVs in the LMC discussed in the text.   The apparent transits of the LBV during outburst are shown as straight blue lines. The LBV/S Dor instability strip is outlined in pink, and the empirical upper luminosity boundary is shown in orange.  Evolutionary tracks for non-rotating models from \citet{2021A&A...652A.137E} are shown in gray for 20 $M_\odot$, 40 $M_\odot$, 60 $M_\odot$, and 85 $M_\odot$.
\label{fig:hrd_lbv}}
\end{figure}

\citet{1994PASP..106.1025H} divided the LBVs into two groups based on their luminosities in quiescence; the classical LBVs with luminosities above the Humphreys-Davidson limit and the less luminous LBVs.  These two groups have different evolutionary histories. The more luminous classical LBVs do not become red supergiants (RSGs)  while the less luminous group can evolve to the RSG stage and may be  in a post-RSG state. Table \ref{tab:lbv} includes three classical LBVs and five in the less luminous group, plus five candidates. The candidates are those that share some characteristics with LBVs but without an observed S Dor eruption. Figure \ref{fig:hrd_lbv} shows the confirmed LBVs on a schematic HR Diagram. 

\vspace{2mm}

Due to the infrequency of the S Dor variability, very few evolved hot stars have been confirmed as LBV/S Dor variables. The confirmed LBVs in Table \ref{tab:lbv} are the same stars that were in \citet{2016ApJ...825...64H} with two additional members, HDE 269582 and HD 269216 from \citet{2017AJ....154...15W}. HD 269582 was listed in \citet{2016ApJ...825...64H} as a candidate LBV.  The entire class of candidate LBVs should be approached  with caution, because there is no good way to estimate what fraction of them are true LBVs. 

\vspace{2mm} 

\begin{deluxetable*}{cclccc}[b]
\tabletypesize{\scriptsize}
\tablewidth{0pt} 
\tablecaption{Luminous Blue Variables\label{tab:lbv}}
\tablehead{
\colhead{RAJ2000} &\colhead{DecJ2000} & \colhead{Name} &  
\colhead{Sp. Type} & \colhead{M$_{Bol}$} & 
\colhead{References\tablenotemark{a}}\\
}
\startdata 
\multicolumn{6}{c}{Classic LBVs}\\
\hline
79.55979&-69.25028&S Dor (HD 35343)&B1&-9.8&1,2,11\\ 
	81.96943&-68.98569&HDE 269582 (Sk -69 142a)\tablenotemark{b}&Ofpe/WN10h&-9.7&9,11\\
84.18208&-69.49650&R127 (HDE 269858)&Ofpe/WN9&-10.5&1,3,11\\ 
\hline
\multicolumn{6}{c}{Less-Luminous LBVs}\\
\hline
75.53073&-71.33703&R71 (HDE 269006)&B5:I&-9.2&5,6,11\\ 
78.37821&-69.53989&HDE 269216 (Sk -69 75)&B7:Iae&-9.1&11\\
79.48363&-69.26869&R85 (HDE 269321)&B5Iae&-8.5&8\\
82.71446&-69.04963&R110 (HDE 269662)&B6 I&-8.9&7\\
84.71508&-69.13531&R143& B9 - A  & -8.6&4,11,13\\
\hline
\multicolumn{6}{c}{Candidate LBV}\\
\hline
82.85635&-69.09404&S119 (HDE 269687)&WN11h&-9.8&9\\
	82.96783&-68.54413&R116\tablenotemark{c} (HDE 269700)& B1.5aeq & -10.6 & 4, 14\\
83.92683&-69.67734&HD 37836&B0e(q)&-11.4&15\\
84.99476&-69.734465&R149 (Sk -69 257)& O8.5 II((f)))& -9.7&12\\  
85.43607&-69.58747&Sk -69 279&O9f&-9.7&10\\
86.46641&-67.24054&S61 (Sk -67 266)&WN11h&-9.7&9\\
\enddata
\tablenotetext{a}{(1) \citet{1986AA...154..243S}, (2) \citet{1985AA...153..168L}, (3) \citet{1983AA...127...49S}, (4) \citet{1960MNRAS.121..337F}, (5) \citet{1981AA....99..351W}, (6) \citet{2013AA...555A.116M}, (7) \citet{1990AA...228..379S}, (8) \citet{2000AJ....119.2214M}, (9) \citet{1997AA...320..500C}, (10) \citet{1986AJ.....92...48C}, (11) \citet{2017AJ....154...15W}, (12) \citet{2009AJ....138..510F}, (13) \citet{2019AA...626A.126A}, (14) \citet{1972AAS....6..249A}}, (15) \citet{1984ApJS...55....1S}
\tablenotetext{b}{Included as a candidate LBV in Humphreys et al 2016. It is assigned a luminosity both higher \citep{1997AA...320..500C} and lower \citep{2023MNRAS.521..585C} than the Humphreys-Davidson Limit ($M_{bol} < -9.7$) with the discrepancy likely due to a difference in adopted extinction.  We have included 
it here as a classic high luminosity LBV because \citet{2017AJ....154...15W} noted strong spectroscopic similarities with LBV R127, and argued it is a similar case. }
\tablenotetext{c}{\citet{2021AA...655A..98A} included R116 as a confirmed LBV. \citet{2001AA...366..508V} listed it as an LBV based on 19th century eye estimates that R116 was about half a magnitude brighter than modern photometry from different sources since 1956 \citep{1999AA...349..537V}. It has not shown any S Dor variability or any significant variability in 70 years. We include it as a candidate LBV.} 
\end{deluxetable*}

R143 has a checkered history. It is noted here because it is pertinent to discussion in later sections. It was formerly considered a classical LBV, but its identification in quiescence with HDE 269929, a late O-type star, by \citet{1993ApJ...409..770P} was due to a misidentification \citep{2019AA...626A.126A}. However, the most recent spectroscopic observations summarized by \citet{2017AJ....154...15W} show variation between late B and A-type. Based on its earlier F-type spectrum, observed from 1952 to 1985 \citep{1960MNRAS.121..337F,1985AA...153..235M}, R143's maximum light or LBV eruption lasted more than 30 yrs. \citet{2019AA...626A.126A} report a temperature of 8500 K and M$_{Bol}$ of -8.6 mag in its current quiescent state.  Thus during its high mass loss F-type supergiant phase, R143's bolometric luminosity, -9.4 to -9.1 mag,  may have increased compared to its luminosity in quiescence.  In that respect, it may be like R71 which is currently experiencing an apparent increase in luminosity during its LBV eruption \citep{2013A&A...555A.116M,2017AJ....154...15W}. R143 is included in Table \ref{tab:lbv} as a less luminous LBV. 

\vspace{2mm}

In quiescence or their normal hot star state LBVs are often late-WN type stars or are classified as mid to late B-type supergiants with emission lines. Consequently, many hot, emission line stars may be confused with LBVs.  The most common are the B[e]sgs which spectroscopically resemble LBVs. One of the distinguishing characteristics of B[e]sgs is the presence of warm dust in the near-infrared not observed in LBVs \citep{2019Galax...7...83K,1998A&A...340..117L}.  The B[e] supergiants are included here in Table \ref{tab:besg}.

\vspace{2mm}

Several published lists of LBVs and candidates include numerous B[e]sgs.   For example, \citet{2018RNAAS...2..121R} included six B[e]sg in the LMC  plus  $\alpha$ Cyg variables R74, R78, and HD 269604 \citep{1999AA...349..537V}.  \citet{2018AJ....156..294A} listed three known B[e]sg and two $\alpha$ Cyg variables. \citet{2021AA...655A..98A} separated the B[e]sg, but included two non-LBVs, R81 and R99, and two $\alpha$ Cyg variables as candidates. The kinematic study by \citet{2022MNRAS.516.2142A} also listed five non-LBV stars and a B[e]sg.  The non-LBVs in \citet{2022MNRAS.516.2142A} were previously discussed in \citet{2016ApJ...825...64H}. They are;  R84 a composite B0 Ia + M4 Ia, Sk -69$^{\circ}$ 271 classified B4I/III, and R99, a peculiar Ofpe/WN9 star but not considered an LBV. The description of these stars is not repeated here. R81 and MWC 112, however, require some discussion. 

\vspace{2mm}

{\it R81 (= HDE 269128 = Hen S86)} is not an LBV or candidate. It is a known eclipsing binary \citep{1987A&A...184..193S,2002A&A...389..931T}.  R81's spectroscopic and photometric variability are due to its orbital motion. The primary is an early B-type supergiant, but the secondary is not observed and assumed to be a small, cooler star. The pair is an interacting binary with a short orbital period, and \citet{2002A&A...389..931T} hypothesize that the secondary may be surrounded by material accreted from the primary. They find that the spectroscopic variations are synchronized with the orbit.  \citet{2022MNRAS.516.2142A} refer to the LBV, presumably the hot primary, but no S Dor variability has been reported.  \citet{2021AA...655A..98A} also list it as a candidate LBV.

\citet{2016ApJ...825...64H} described {\it MWC 112} as an F5 Ia supergiant and identified it with Sk -69$^{\circ}$ 147 based on its entry in SIMBAD. A closer examination by \citet{2023AJ....166...50H} showed that Sk -69$^{\circ}$ 147 is indeed a normal F-type supergiant. However, MWC 112 is HDE 268582 = Sk -69$^{\circ}$ 142a \citep{1996A&A...308..763V}, now recognized as an LBV (Table \ref{tab:lbv}). \citet{2022MNRAS.516.2142A} list MWC 112 and Sk -69$^{\circ}$ 142a  separately in their Table 1. It is unclear which star they observed as MWC 112. The entry in SIMBAD has been corrected. 

\begin{deluxetable*}{cclcccc}[t]
\tabletypesize{\scriptsize}
\tablewidth{0pt} 
\tablecaption{B[e] Supergiants\label{tab:besg}}
\tablehead{
\colhead{RAJ2000} &\colhead{DecJ2000} & \colhead{Name} &  
\colhead{Sp. Type} & \colhead{M$_{Bol}$} &
\colhead{Status\tablenotemark{a}}&
\colhead{References\tablenotemark{b}}\\
}
\startdata 
\multicolumn{7}{c}{Early}\\
\hline
74.40317&-67.79371&LHA 120-S 12&B0.5Ie&-8.7&Confirmed&1\\
78.47073&-67.44856&HD 34664&B0/0.5I[e]&-10&Established&2\\
81.82433&-66.36823&LHA 120-S 35&Hot&-8.2&Established&3\\
82.09463&-69.14219&HDE 269599&B1&-10.4&Established&5\\
84.10776&-69.38218&HD 37974&B0.5e&-10.5&Established&5\\
85.05554&-69.37958&HD 38489&B0&-9.8&Establised&6\\
85.43232&-69.62732&LHA 120-S 137&BIaePCyg&-5.8&Established&7\\
86.37289&-68.19608&LHA 120-S 59&B0.5IIIe&-5.2&Established&4\\
\hline
\multicolumn{7}{c}{Mid}\\
\hline
78.41205&-69.35230&HDE 269217&B5/6Ia[e]&-8.8&Established&10\\
\hline
\multicolumn{7}{c}{Late}\\
\hline
73.68104&-70.35762&ARDB 54&A0-1I&-6.1&Confirmed&12\\
74.19606&-69.84095&HDE 268835&B8[e]&-8.9&Established&1\\
79.13247&-68.36926&LHA 120-S 93&A0:I:&-6.7&Established&11\\
\enddata
\tablenotemark{a}{Classification status as assigned by \citet{2019Galax...7...83K}}
\tablenotetext{b}{(1) \citet{1986AA...163..119Z}, (2) \citet{1996ApJ...465..231O}, (3) \citet{1995AA...302..409G}, (4) \citet{2012MNRAS.425..355R}, (5) \citet{1984ApJS...55....1S}, (6) \citet{1983ApJ...273..177S}, (7) \citet{2009AJ....138..510F}, (8) \citet{1972AAS....6..249A}, (9) \citet{1978AAS...31..243R}, (10) \citet{2001AA...368..160C}, (11) \citet{1970CoTol..89.....S}, (12) \citet{2019MNRAS.488.1090C}}
\end{deluxetable*}

\subsection{B[e] Supergiants} \label{sec:besg}

The evolutionary state of the B[e]sg is uncertain. Like the LBVs, they are massive post-main sequence stars. Some authors have argued for a post-RSG or post-yellow/warm supergiant stage \citep{2010AJ....139.1993K,2012MNRAS.423..284A} and others for a pre-RSG state \citep{2013A&A...558A..17O,2014Msngr.157...50D}. In addition to their emission line spectra, the B[e]sg are distinguished by their near-infrared excess radiation and dusty envelopes which appear to be concentrated in the equatorial region and may obscure much of the star as shown by the SEDs of several in M31 and M33 \citep{2017ApJ...844...40H}.  There is increasing evidence from studies of individual B[e]sg that they are close, interacting binaries with rapid rotation driven mass loss in the equatorial region \citep{2014Msngr.157...50D}. 

\vspace{2mm} 

\begin{figure}[b]
\epsscale{0.9}
\plotone{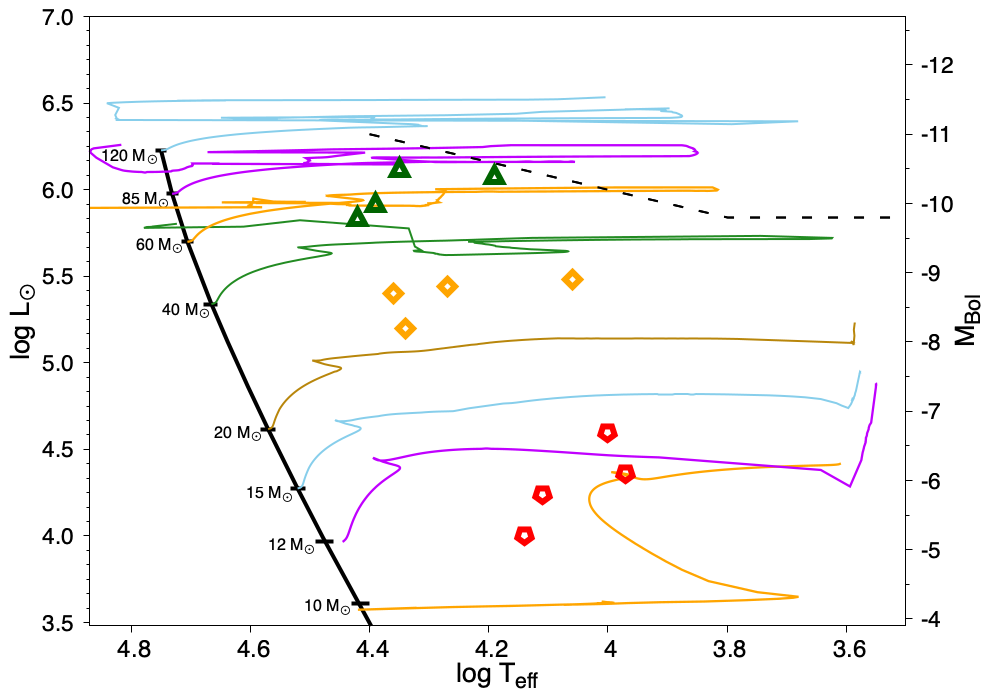}
\caption{An HR diagram of the B[e]sg in the LMC sorted into three luminosity groups: high luminosity B[e]sg (green triangles, $M_{Bol} < -9.7$), mid-luminosity B[e]sg (orange diamonds, $-9.7 \leq M_{Bol} < -7.0$), and low luminosity B[e]sg (red pentagons, $M_{Bol} \geq -7.0$).  The ZAMS and evolutionary tracks are adopted from \citet{2021A&A...652A.137E} Z= 0.006 non-rotating models which is consistent with the range of spectroscopic abundances measured for luminous stars in the LMC \citep{2022A&A...658A..29R,2019PASP..131b4101A,2008AJ....136..375M}. 
\label{fig:hrd_besg}}
\end{figure}

The list of B[e]sg in Table \ref{tab:besg} is primarily from \citet{2019Galax...7...83K}. In the LMC the B[e]sg are found over a range of spectral types and luminosity from early B-type supergiants to late B and early A-type stars. This is similar to what \citet{2017ApJ...844...40H}  found in M33.  Figure \ref{fig:hrd_besg} shows an HRD diagram with the B[e]sg.  Their luminosity distribution clearly suggests a wide range in initial masses for the progenitors of the B[e]sg class with potentially three recognizable groupings by luminosity with four in each group; highest luminosity B[e]sg ($M_{Bol} < -9.7$), mid-luminosity ($-9.7 \leq M_{Bol} < -8$), and the lowest luminosity ($M_{Bol} \geq -7$). 

\subsection{Wolf-Rayet Stars} \label{sec:wr}

\begin{figure}[t]
\epsscale{.9}
\plotone{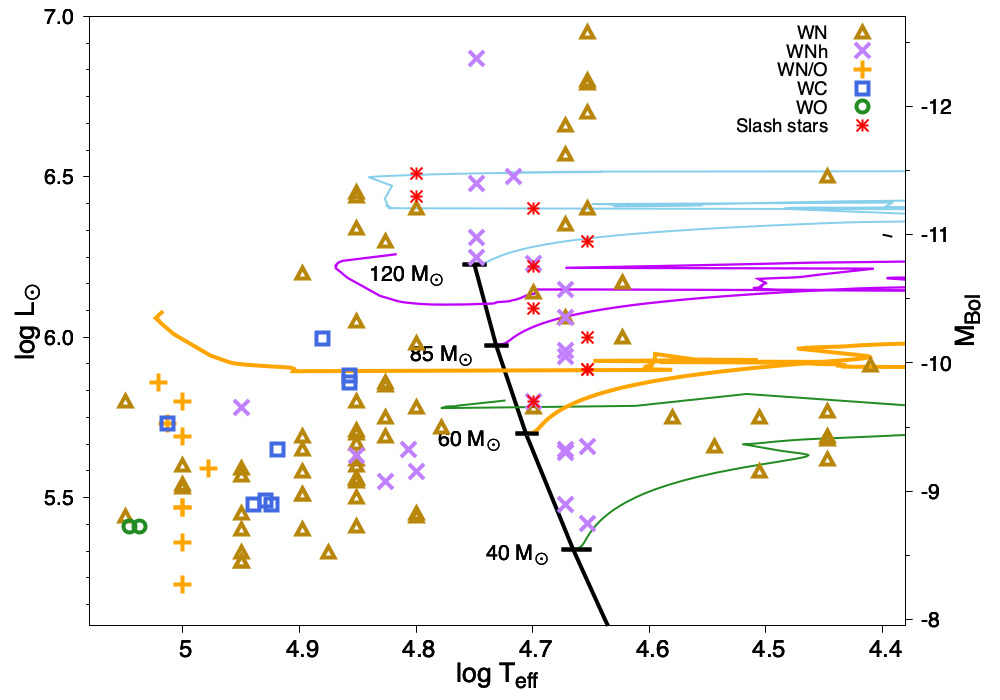}
\caption{An HR diagram of the 126 of 152 WR in the LMC which have published temperature and luminosity \citep{1998AAS..130..527T,2000AJ....119.2214M,2014AA...570A..38B,2014AA...565A..27H,2014MNRAS.442..929G,2017ApJ...841...20N,2022ApJ...931..157A}.  WN type are gold triangles.  WNh type are purple X's. WN3/O3 are orange crosses, WC type are blue squares, WO type are green circles.  O2-3If*/WN "slash stars" are red asterisk. The ZAMS and evolutionary tracks are adopted from \citet{2021A&A...652A.137E} Z= 0.006 non-rotating models which is consistent with the range of spectroscopic abundances measured for luminous stars in the LMC \citep{2022A&A...658A..29R,2019PASP..131b4101A,2008AJ....136..375M} 
\label{fig:hrd_wr}}
\end{figure}

We sorted the list of LMC Wolf-Rayet (WR) stars from \citet{2018ApJ...863..181N} into six classifications:  WN, WNh, WN3/O3, Of/WN slash-stars, WC, and WO.  The WN and WNh stars are separated into high ($M_{Bol} < -9.7$) and less luminous ($-9.7 \leq M_{Bol} > -8.0$) groups. The slash-stars are all high lumninosity.  The WC, WO, and WN3/O3 are predominantly lower-luminosity.  Appendix \ref{wrlisting} includes the details for each WR sample in our study. The WR of different types which have published luminosities and temperatures (127 out of 152) are shown on the HR Diagram in Figure \ref{fig:hrd_wr}.

\section{Spatial Distribution and Nearest Neighbor Analysis} \label{analysis}

Most of the stars discussed in this paper are post-main sequence. This is clear from their spectroscopic properties; temperatures, luminosities, and positions on the HR Diagram. We should not expect to find them closely associated with main sequence stars. Their appropriate comparison population should be the evolved stars that share the same range of initial masses and ages. On the HR Diagram, we use luminosity as a substitute for mass and temperature or spectral type for age or evolutionary state. In this section we focus on  the spatial distribution of each group.

\vspace{2mm}

As they age, stars are expected to move from their birthplace. The stars they were born with will do likewise, so we expect an increasingly more dispersed comparison population as the stars age and evolve. For example, the B[e]sgs divide into three groups on the HR Diagram (Figure \ref{fig:hrd_besg}).  Based on their positions on the HR Diagram, we expect the most luminous to be associated with mid to late O-type stars, while the least luminous group shares space on the HR diagram with evolved 12 to 15 M$_{\odot}$ stars that may be evolving to the red supergiant stage\footnote{The mass estimates in
this paper are based on the luminosities of the stars in comparison 
with the evolutionary tracks.}. The more massive stars that they may have formed with will have already become RSGs or even supernovae.  

\vspace{2mm} 

In the following discussion, we have identified reference populations in the LMC based on their positions on the HR Diagram in Figure \ref{fig:hrd}. The reference samples are initially divided into three groups based on luminosity, a substitute for initial mass: (1) the most luminous, above the H-D limit at M$_{Bol} < -9.7$ (Log (L/L$_{\odot}) > 5.7$), (2) less luminous stars with  M$_{Bol}$ between -8.0 and -9.7, and (3) a limited group with luminosities M$_{Bol}$ between -5.0 and -7.0, chosen as a reference sample for the lowest luminosity B[e]sg.  Further subdivision is based on spectral types as an indicator of evolutionary state.  The reference samples are summarized in Table \ref{tab:refsamp}.  Six high luminosity yellow hypergiants (YHG) with circumstellar dust and high mass loss \citep{2023AJ....166...50H}  which may be indicative of post-red supergiant evolution, have been separated from the yellow supergiants. Known and suspected binaries are included based on their spectral types

\vspace{2mm}

The spatial distribution of the reference samples are shown on large maps in Appendix \ref{lmcmaps}. These maps show that the earliest type stars, near the main sequence, are more tightly clustered than those with older evolutionary ages. This demonstrates the importance of comparing our target groups with a reference sample that shares their space on the HR Diagram and are likely to have similar ages.

\begin{deluxetable*}{lccc}
\tabletypesize{\scriptsize}
\tablewidth{0pt} 
\tablecaption{Reference Samples\label{tab:refsamp}}
\tablehead{
\colhead{Name}&\colhead{Number}&
\colhead{Spectral Types}&
\colhead{Median Sep. (pc)\tablenotemark{a}}
}
\startdata 
\multicolumn{4}{c}{High Luminosity Samples ($M_{Bol} < -9.7$)\tablenotemark{c}}\\
\hline
Early O-types&38&O2-O4&41\\
Mid O-types&47&O4.5-O7&24\\
Late O-types&15&O7.5-O9&329\\
Early B (high luminosity)&47&O9.2-B2.5&124\\
\hline
\multicolumn{4}{c}{Less Luminous Samples for LBVs, B[e]sg, and WR ($-8.0 > M_{Bol} \geq -9.7$)\tablenotemark{c}}\\
\hline
Early + Mid O-types&212&O2-O7&14\\
Late O + Early B-types&610&O7.5-B2.5&27\\
Mid + Late B-Types&76&B3-A0&136\\
A Supergiants&9&A1-A8&258\\
Yellow Supergiants (YSG)&22&A9-K4.5&282\\
Yellow Hypergiants (YHG)\tablenotemark{b}&6&A9-K4.5&1367\\
Red Supergiants (RSG)&45&K5 and later&113\\
\hline
\multicolumn{4}{c}{Lowest Luminosity Samples for B[e]sg ($-5 > M_{Bol} \geq -7$)\tablenotemark{c}}\\
\hline 
All O-type + Early B-type&791&O2-B2.5&14\\
Mid B-type&162&B3-B7&138\\
Late B-type&324&B7.5-A0&108\\
A Supergiants&183&A1-A8&125\\
Yellow Supergiants (YSG)&34&A9-K4.5&437\\
Red Supergiants (RSG)&109&K5 and later&109\\
\hline
\enddata
\tablenotetext{a}{The median separation of the stars in the population from each other computed by performing a nearest neighbor analysis of the population relative to itself.}
\tablenotetext{b}{The yellow hypergiants have circumstellar dust and high mass loss rates which may be due to a post-RSG evolutionary state \citep{2023AJ....166...50H}.}
\tablenotetext{c}{While correspondence between luminosity and mass is probably affected by rotation, the highest luminosity range roughly corresponds to main sequence stars above $>60 M_\odot$ and evolved stars above $> 40 M_\odot$.  The middle luminosity range roughly corresponds to main sequence stars between 30--60 $M_\odot$ and evolved stars between 20--40 $M_\odot$.  The lowest luminosity range roughly corresponds to main sequence stars between 12--20 $M_\odot$ and evolved stars between 10--15 $M_\odot$.}
\end{deluxetable*}

While many stars with relatively older evolutionary ages tend to be more dispersed, the 30 Doradus region includes at least five distinct stellar groups ranging in evolutionary ages from ongoing star formation to more than 10 Myr old \citep{1997ApJS..112..457W}.  Thus stars of a more advanced evolutionary age may be found in relatively close projected proximity to much younger stars. The impact this clustering has on the analysis  will be discussed.

\vspace{2mm}

We present a simple form of nearest-neighbor analysis with cumulative distribution plots like those in the statistical Kolmogorov-Smirnov (K-S) tests. In a nearest neighbor analysis, the relative spacing between members of two different populations is only meaningful if both populations share a characteristic spatial-dispersion size-scale. But, several differing size-scales coexist in the LMC. (1) each star formation region has its own age and velocity dispersion,   
(2) the velocity dispersion may depend on stellar mass, 
(3) for some stars the distance from the nearest neighbor, $D_1$,  is related to the separation between two star formation centers, 
and (4) the LMC has its own global size scale. The multiple spatial scales complicate the use of the standard K-S test.  For these reasons, we measure the dispersion of each reference population relative to itself.  Older populations with more time to disperse from their point of origin will tend to have a greater median distance between members (Table \ref{tab:refsamp}).  The more alike the cumulative distributions are the more likely the sample is drawn from that population.

\vspace{2mm}

We measure the distance from the nearest neighbor in the reference populations for each target in our LBV, B[e]sg and WR samples, and compare the results on a cumulative distribution plot.  However, caution is called for when using statistics to characterize small samples such as our LBV and B[e]sg targets with only a few stars.  Samples smaller than N $<$ 20 cannot yield K-S test results with strong significance \citep{1986nras.book.....P} and samples smaller than N $<$ 8 cannot be analyzed with even a one sigma significance.  See \citet{2016arXiv160802007D} for a detailed discussion about how sampling can influence the nearest neighbor analysis. Hence we have not included any formal K-S tests. We show the cumulative distributions here for visual inspection and for comparison with previous work \citep{2015MNRAS.447..598S,2016ApJ...825...64H, 2016arXiv160802007D,2019MNRAS.489.4378S,2018AJ....156..294A,2025RAA....25d5001K}  and the recent paper by \citet{2024ApJ...976..125D}. 

\vspace{2mm}

\subsection{The Luminous Blue Variables}

\begin{figure}
\plotone{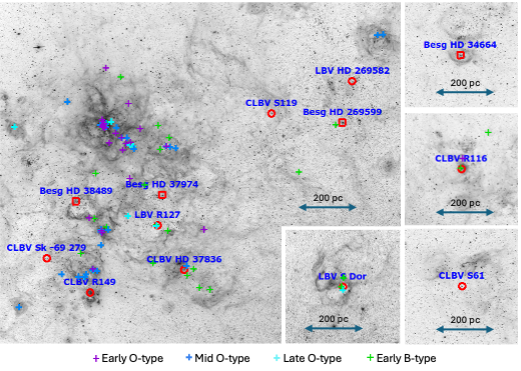}
\caption{The high luminosity ($M_{Bol} < -9.7$) LBVs, candidate LBVs, and the B[e]sg plotted with red circles on H-alpha emission images of the LMC from MCELS \citep{1999IAUS..190...28S}.  The plots also include the reference samples in the matching luminosity range (Table \ref{tab:refsamp}): Early O-types (purple), Mid O-types (blue), Late O-types (cyan), and Early B-types (green).}
\label{fig:LMCMap_hilum_zoom}
\end{figure}

\begin{figure}
\plotone{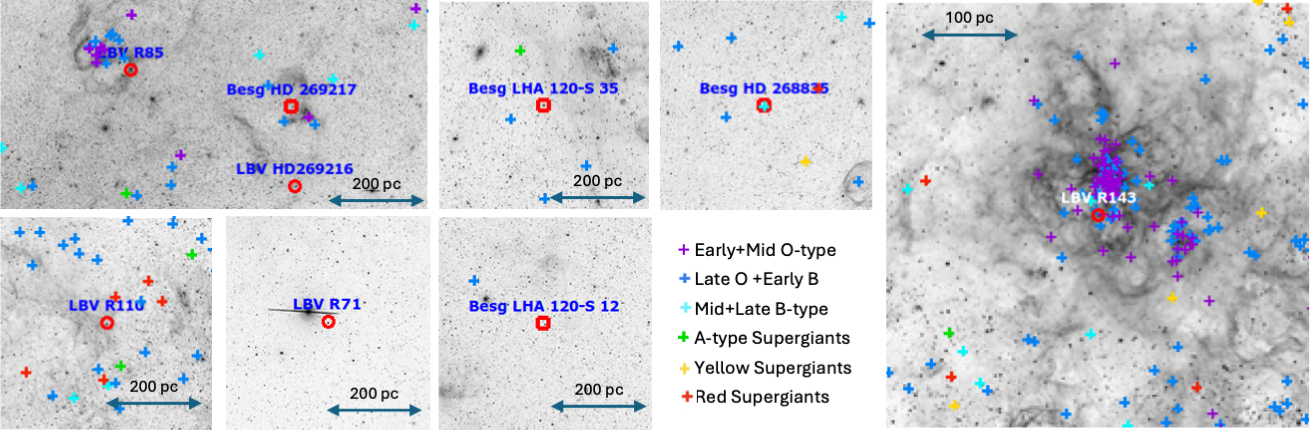}
\caption{The less luminous ($-9,7 \leq M_{Bol} < -8$) LBVs and the B[e]sg plotted with red circles on H-alpha emission images of the LMC from MCELS \citep{1999IAUS..190...28S}.  The plots also include the reference samples in the matching luminosity range (Table \ref{tab:refsamp}): Early + Mid O-types (purple), Late O-types + Early B-types (blue), and Mid + Late B-types (cyan), A-type supergiants (green), yellow supergiants (gold), and red supergiants (red).}
\label{fig:LMCMap_midlum_zoom}
\end{figure}

Figures \ref{fig:LMCMap_hilum_zoom} and \ref{fig:LMCMap_midlum_zoom} show the confirmed and candidate LBVs on an H-alpha image of the LMC with the reference populations defined in Table \ref{tab:refsamp}. The conclusion is clear. Most LMC LBVs are in the line of sight with known H II regions. They are not isolated from other massive stars. Candidate LBV S61 and less luminous LBV R71 are exceptions. Each is more than 350 pc from the nearest star of comparable luminosity in the reference samples.  S61, however, is close to four higher luminosity  WN stars (see Figure \ref{fig:wr_hi_lum_Map}). Candidate LBV S119 is also somewhat isolated, 220 pc from its nearest neighbor of comparable luminosity, but it is also associated with a region northwest of the Tarantula nebula which includes H-alpha emission and many stars with $-8.0 > M_{Bol} \geq -9.7$ including the less luminous LBV R110.  Thus S119 ($M_{Bol} = -9.8$) may be evolved from a less luminous progenitor.

\subsubsection{The Classical (High Luminosity) LBVs\label{analysisclassiclbv}}

Figure \ref{fig:LMCMap_hilum_zoom} shows the S Dor region and  images centered on the classical LBVs together with the stars from the high luminosity ($M_{Bol} < -9.7$) reference populations in the same projected area. The candidate LBVs are included because they all fall into the high luminosity group. The identification of any H II region or cluster associated with the targets is given in Table \ref{tab:high_lum}. S Dor and HDE 269582 are in clusters between 1-6 Myr old \citep{1985AA...152..427C,2007ApJ...655..179W,2012ApJ...751..122P,2016MNRAS.457.2151A}. R127 is close to main sequence early O-type stars as well as a 3-6 Myr old cluster containing evolved late O-type and early B-type supergiants \citep{2003AA...400..923H}. The common factor between all three classic LBV is their close proximity to evolved late O-type and early B-type post-main sequence stars. The candidate-LBVs are also associated with evolved late O-type and early B-type supergiants. R149 is in a star cluster with an estimated age of 4-6 Myr \citep{1985AA...152..427C}.

\begin{figure}
\epsscale{0.7}
\plotone{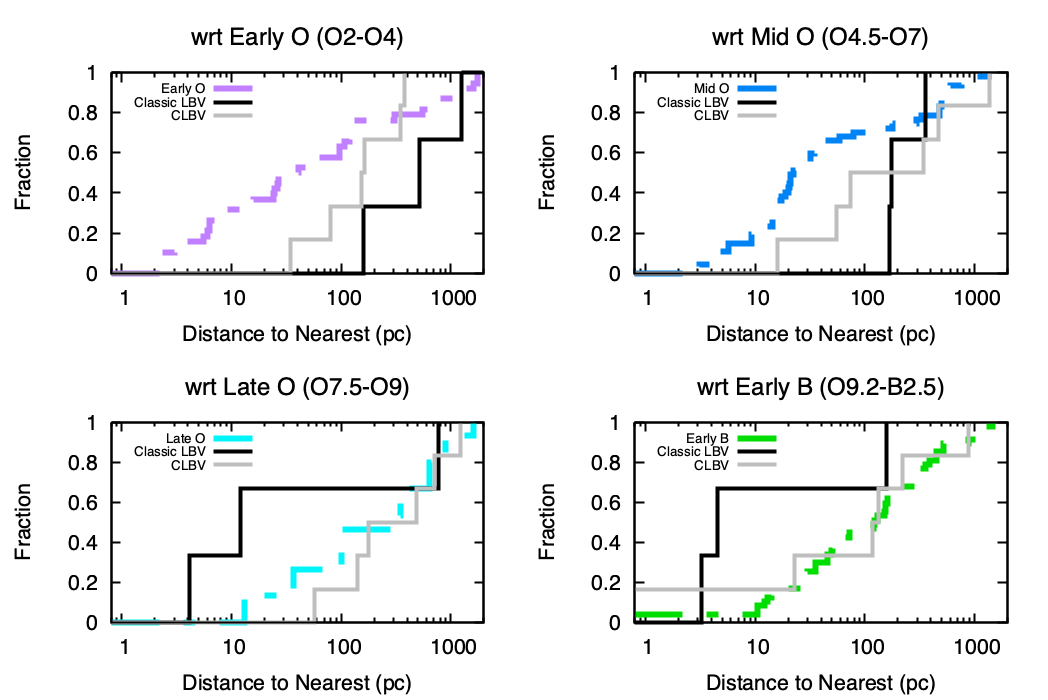}
	\caption{Cumulative distribution plots of the nearest neighbors for classic LBV (black) and candidate LBV (gray) plotted relative to high luminosity $M_{Bol} < -9.7$ reference samples (colored dotted line).}
\label{fig:CumPlotsClassicLBV}
\end{figure}

The results of the nearest neighbor analysis  (Figure \ref{fig:CumPlotsClassicLBV}) show that the classic LBVs and candidate LBVs best match the distributions of the late O and early B-type supergiants in the same luminosity range. The effect of the small sample size for the classic LBVs is clear. Even though there are only three classic LBVs, they are  more closely associated with evolved late-O and early B-type supergiants with evolutionary ages of  2-4 Myr \citep{2021A&A...652A.137E} rather than with main sequence stars of the same luminosity. The spectral type of the nearest star to each target with similar luminosity (Table \ref{tab:high_lum}) reinforces that result.

\begin{figure}[bt!]
\epsscale{0.63}
\plotone{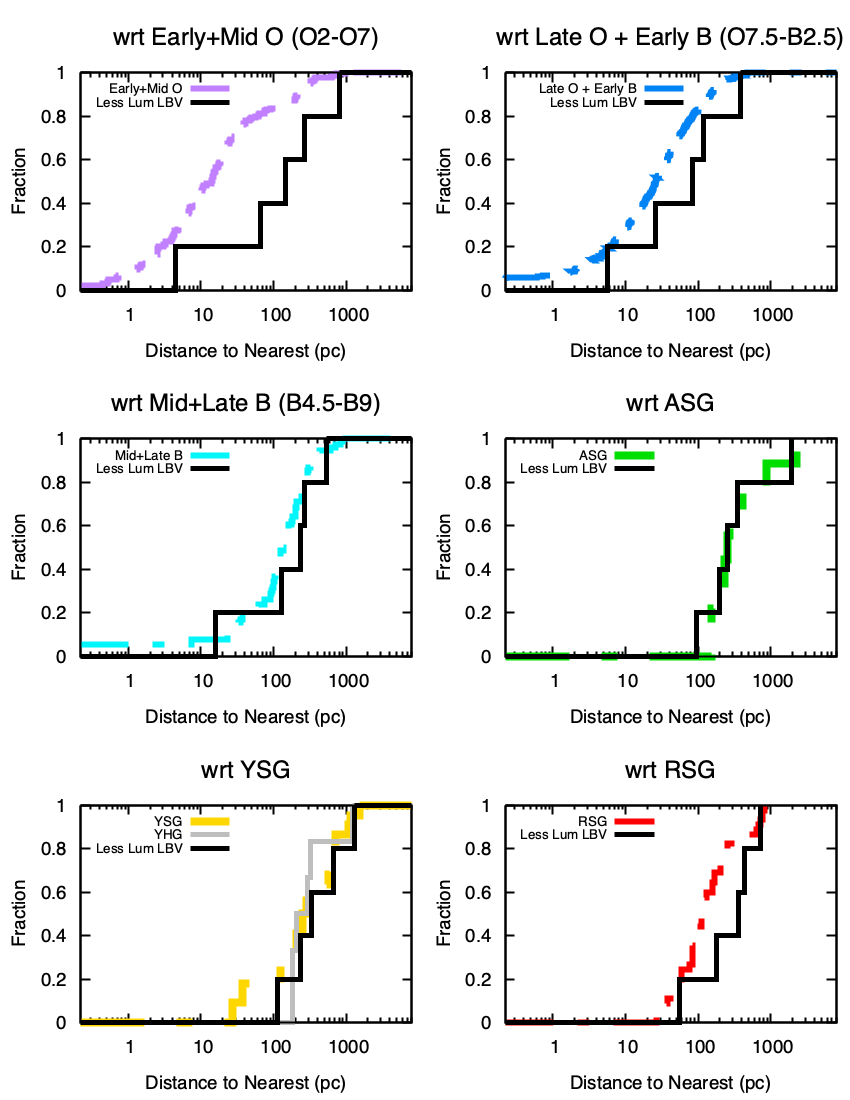}
\caption{Cumulative distribution plots of the nearest neighbors for the less luminous LBVs (black) plotted relative to mid luminosity $-8.0 > M_{Bol} \geq -9.7$ reference samples (colored dotted line).  The YHG sample is also plotted relative to the YSG as a gray line in that panel.}
\label{fig:lesslumLBV}
\end{figure}

\subsubsection{Less Luminous LBVs}\label{analysis-lowlumlbv}
Figure \ref{fig:LMCMap_midlum_zoom} shows images centered on the less luminous LBVs with stars of similar luminosity ($-8.0 > M_{Bol} \geq -9.7$).  In this luminosity range the main sequence corresponds to early and mid O-type stars. As with the highest luminosity LBVs, a few are close to main sequence stars of similar luminosity while others are more isolated. All but one are found close to evolved B-type stars. R71  is isolated from every type of star in the same luminosity range.  R85 is in the southwest part of the large spiral shaped emission nebula LHA-120 N119 and OB association NGC 1910 (LH 41) which includes at its heart several evolved supergiants including LBV S Dor. R110 resides in an area just northwest of the Tarantula Nebula which includes relatively few high luminosity stars (Figure \ref{fig:LMCMap_hilum}) but many stars of comparable luminosity including several RSG, one WN, and one WC, implying a mix of evolved stars more than 3 Myr old.

\movetabledown=2.in
\begin{rotatetable}    
\begin{deluxetable*}{lrrrrrrrrrrrrcclc}
\tabletypesize{\scriptsize}
\tablewidth{0pt} 
\tablecaption{Distance to Nearest Neighbors for with Respect to Reference Samples $M_{Bol} < -9.7$\label{tab:high_lum}}
\tablehead{
&\multicolumn{3}{c}{Early O-type\tablenotemark{a}}
&\multicolumn{3}{c}{Mid O-type\tablenotemark{a}}
&\multicolumn{3}{c}{Late O-type\tablenotemark{a}}
&\multicolumn{3}{c}{Early B-type\tablenotemark{a}}
\\
&
\colhead{D1}&\colhead{D2}&\colhead{D3}&
\colhead{D1}&\colhead{D2}&\colhead{D3}&
\colhead{D1}&\colhead{D2}&\colhead{D3}&
\colhead{D1}&\colhead{D2}&\colhead{D3}&
\multicolumn{2}{c}{Nearest\tablenotemark{c}}&\multicolumn{2}{c}{In Line of Sight\tablenotemark{b}}\\
\colhead{Star}&
\colhead{(pc)}&\colhead{(pc)}&\colhead{(pc)}&
\colhead{(pc)}&\colhead{(pc)}&\colhead{(pc)}&
\colhead{(pc)}&\colhead{(pc)}&\colhead{(pc)}&
\colhead{(pc)}&\colhead{(pc)}&\colhead{(pc)}&
\colhead{SpType}&D(pc)&\colhead{Name}&\colhead{Age (Myr)}
}
\startdata 
\multicolumn{16}{c}{Classic LBV}\\
S Dor (HD 35343)&1236&1297&1383&359&746&756&12&360&1198&3&28&282&O9.5&3&H88-267,  LHA-120 N119, DEM L132a&1-6\tablenotemark{1}\\
HDE 269582 (Sk -69 142a)&520&669&714&171&192&193&776&786&796&158&497&547&B0&158&LH 61, NGC 1983&1-6\tablenotemark{2}\\
R 127 (HDE 269858)&159&184&189&177&190&642&4&103&268&5&42&130&O8.5&4&NGC 2055&3-6\tablenotemark{3}\\
\hline
\multicolumn{16}{c}{Candidate LBV}\\
S119 (HD 269687)&373&447&456&346&363&376&482&493&544&220&223&357&B0&220&BSDL 2141&-\\
R 116 (HD 269700)&35&598&617&472&490&655&704&720&724&0&153&576&B1.5&0&LHA 120-N 148&-\\
HD 37836&152&302&307&16&291&292&175&261&308&23&33&81&O6&16&LHA 120-N 154&-\\
R 149 (Sk -69 257)&80&205&212&55&62&65&57&76&289&135&227&236&O6&55&NGC 2083, 	LHA 120-N 159&4-6\tablenotemark{4}\\
Sk -69 279&162&197&206&73&125&143&142&161&310&117&208&209&B1&117&[GKK2003] O161&-\\
S61 (Sk -67 266)&344&891&941&1369&1604&1627&1221&1623&1732&875&920&922&O3&334&[KDS99] GS 94, DEM L 308&-\\
\hline
\multicolumn{16}{c}{High Luminosity B[e]sg}\\
HD 34664 (B0/0.5I[e])&1447&1506&1520&522&814&839&766&1549&1579&299&1038&1097&B1&299&NGC 1871, LHA 120-N 30&-\\
HD 269599 (B1)&596&617&683&316&327&571&712&724&724&23&224&588&B0&23&NGC 1994 (BSDL 1890)&6-9\tablenotemark{5}\\
HD 37974 (B0.5e)&123&164&185&164&164&164&105&135&191&67&105&125&B0&67&NGC 2050 (BSDL 2554)&-\\
HD 38489 (B0)&71&131&139&128&132&144&183&240&256&86&86&116&O4&71&NGC 2081&3-4\tablenotemark{4}\\
\enddata
\tablenotetext{a}{Reference samples include stars in the same luminosity range ($M_{bol} < -9.7$). D1, D2, and D3 are distances to the first, second, and third nearest stars in the reference sample in parsecs.}
\tablenotetext{b}{Determined using \citet{2008MNRAS.389..678B}.}
\tablenotetext{c}{Spectral type and distance in parsecs of the nearest other star in the same luminosity range.}
\tablenotetext{1}{\citet{1985AA...152..427C}, \citet{2003AJ....126.1836H}, \citet{2007ApJ...655..179W}, and \citet{2012ApJ...751..122P}}
\tablenotetext{2}{\citet{2003AJ....126.1836H}, \citet{2007ApJ...655..179W}, \citet{2012ApJ...751..122P}, and \citet{2016MNRAS.457.2151A}}
\tablenotetext{3}{\citet{2003AA...400..923H}}
\tablenotetext{4}{\citet{1985AA...152..427C}}
\tablenotetext{5}{\citet{1985AA...152..427C}, \citet{1996AA...306..125C}, and \citet{2007ApJ...655..179W}}
\end{deluxetable*}
\end{rotatetable}

\movetabledown=2in
\begin{rotatetable}  
\begin{deluxetable*}{lrrrrrrrrrrrrrclcl}
\tabletypesize{\scriptsize}
\tablewidth{0pt} 
\tablecaption{Distance to Nearest Neighbors with Respect to Reference Samples $-8.0 > M_{Bol} \geq -9.7$ \label{tab:low_lum}}
\tablehead{
&\multicolumn{3}{c}{Late O + Early B\tablenotemark{a}}
&\multicolumn{3}{c}{Mid + Late B\tablenotemark{a}}
&\multicolumn{3}{c}{ASG\tablenotemark{a}}
&\multicolumn{3}{c}{YSG\tablenotemark{a}}
&&\\
&
\colhead{D1}&\colhead{D2}&\colhead{D3}&
\colhead{D1}&\colhead{D2}&\colhead{D3}&
\colhead{D1}&\colhead{D2}&\colhead{D3}&
\colhead{D1}&\colhead{D2}&\colhead{D3}&
\multicolumn{2}{c}{Nearest\tablenotemark{c}}&\multicolumn{2}{c}{In Line of Sight\tablenotemark{b}}\\
\colhead{Star}&
\colhead{(pc)}&\colhead{(pc)}&\colhead{(pc)}&
\colhead{(pc)}&\colhead{(pc)}&\colhead{(pc)}&
\colhead{(pc)}&\colhead{(pc)}&\colhead{(pc)}&
\colhead{(pc)}&\colhead{(pc)}&\colhead{(pc)}&
\colhead{SpType}&\colhead{D(pc)}&\colhead{Name}&\colhead{Age (Myr)}
}
\startdata 
\multicolumn{17}{c}{Lower Luminosity LBV}\\
R 71 (HDE 269006)&388&399&408&530&587&974&1912&2755&2758&1288&1595&1842&B0&388&-&-\\
HDE 269216 (Sk -69 75)&118&134&134&234&281&411&346&1159&1277&336&995&1188&O8.5&118&LH 39, BSDL 940&-\\
R85 (HDE 269321)&26&50&64&265&311&333&255&755&881&681& 739&799&B0.5&26&NGC 1910, LHA-120 N 119, DEM L132b&-\\
R 110 (HDE 269662)&83&96&99&126&168&199&94&225&370&237&249&324&RSG&56&-&-\\
R 143&6&12&14&16&16&17&196&351&514&115&167&246&O5&6&30 Dor, see text &2-3\tablenotemark{2}\\
\hline
\multicolumn{17}{c}{Mid Luminosity B[e]sg}\\
HD 269217 (B5/6Ia[e])&35&61&64&101&126&302&1620&2401&2530&142&567&584&O8&35&LH 35, LHA 120-N 113&6-7\tablenotemark{2}\\
LHA 120-S 12 (B0.5Ie)&168&429&431&578&764&782&2288&2572&2731&611&1181&1270&B0.5&168&-&-\\
HD 268835 (B8[e])&83&156&216&0&243&512&384&1083&1213&409&994&1134&B8&0&-&-\\
LHA 120-S 35 (Hot)&73&154&184&400&658&695&121&2188&2216&741&901&912&B0.5&73&KMHK 915 = [SL63] 482&10-30\tablenotemark{3}\\
\enddata
\tablenotetext{a}{Reference samples include stars in the same luminosity range ($-8.0 > M_{Bol} \geq -9.7$). D1, D2, and D3 are distances to the first, second, and third nearest stars in the reference sample in parsecs.}
\tablenotetext{b}{Determined using \citet{2008MNRAS.389..678B}.}
\tablenotetext{c}{Spectral type and distance in parsecs to the nearest other star in the same luminosity range.}
\tablenotetext{1}{\citet{1997ApJS..112..457W}}
\tablenotetext{2}{\citet{1985AA...152..427C}}
\tablenotetext{3}{\citet{2005AA...442..597V} and \citet{2018AA...612A.113T}}
\end{deluxetable*}
\end{rotatetable} 

R143, now considered a less luminous LBV, (see section \ref{sec:lbv}) is in the 30 Dor region (Figure \ref{fig:LMCMap_hilum_zoom}) surrounded by a mixed population of evolved stars ranging from O to mid B-type supergiants with a likely age of 2-5 Myr \citep{1997ApJS..112..457W}. The closest stars to R143 do not have published spectral types. We use their published photometry with the Q-method to estimate their intrinsic colors and visual extinctions. The A$_{v}$ for R143 \citep{2019AA...626A.126A} is less than for the probable O-type stars and more like that for the candidate B-type supergiants. R143 is therefore likely projected in front of the younger O-type stars, and associated with the B-type supergiants. 

\vspace{2mm}

Figure \ref{fig:lesslumLBV} and Table \ref{tab:low_lum} show the results of the nearest-neighbor analysis of the less luminous LBVs compared to reference samples with $-8.0 > M_{Bol} \geq -9.7$.  The less luminous LBVs have a distribution consistent with the evolved A-type supergiants (ASG) and yellow-supergiants (YSG) of the same luminosity class.  Note also their similarity with the yellow hypergiants which may be post-RSGs.
\vspace{2mm}

The less luminous LBVs range in luminosity from $-9.7 \leq M_{Bol} \leq -8.5$ corresponding to evolved stars with masses between 30 -- 40 $M_{\odot}$ and ages from 4.5 - 6 Myr \citep{2021A&A...652A.137E}. The evolutionary tracks,  with and without rotation, show that the late B-type stars and A supergiants in this luminosity/mass range may spend roughly twice as much time on post-RSG blue evolutionary loops as they do during the post main sequence first crossing.  Both evolutionary states occupy the same temperature and luminosity range so this analysis cannot distinguish between them.

\subsection{B[e]sg}\label{analysis-besg}
The HR-Diagram of the B[e]sg (Figure \ref{fig:hrd_besg}) has three clear groupings by luminosity with four in each group; highest luminosity B[e]sg ($M_{Bol} < -9.7$), mid-luminosity ($-9.7 \leq M_{Bol} < -8$), and the lowest luminosity ($M_{Bol} \geq -7$). 

\vspace{2mm}

\begin{figure}[ht!]
\epsscale{.7}
\plotone{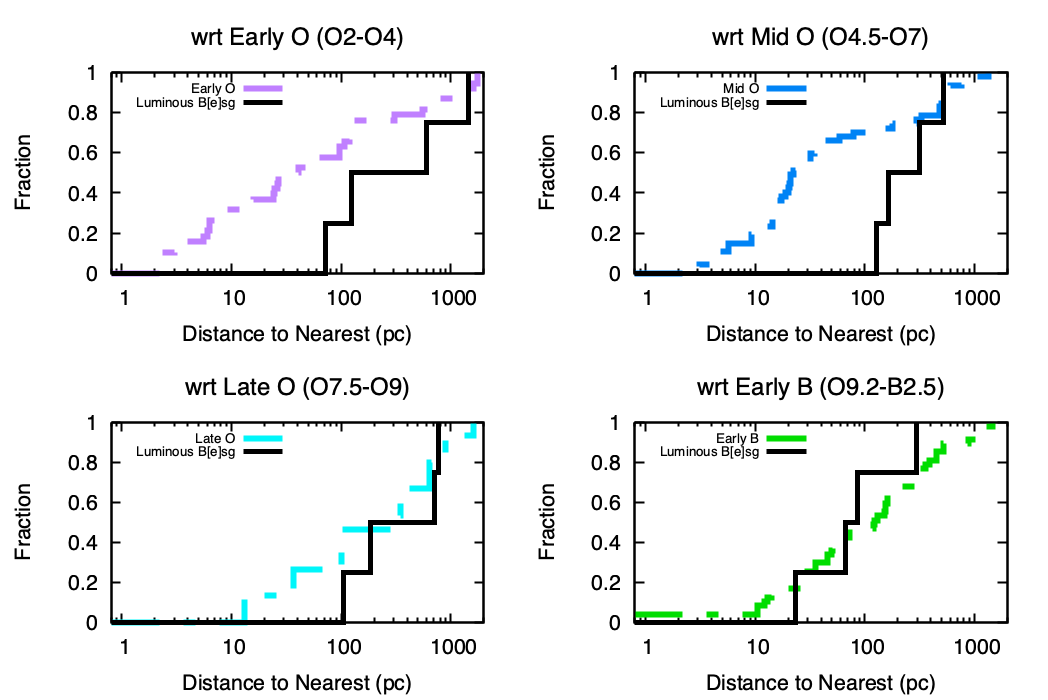}
	\caption{Cumulative distribution plots of the nearest neighbors for the highest luminosity B[e]sg (black) plotted relative to the high luminosity $M_{Bol} < -9.7$ reference samples (colored dotted line).}
\label{fig:classicBesg}
\end{figure}

The highest luminosity B[e]sg ($M_{Bol} < -9.7$) have spectral types between  B0 -- B1.  Plots of the area around each show that all four are in proximity to evolved early B-type stars (Figure \ref{fig:LMCMap_hilum_zoom}). HD 269599 is close to the center of NGC 1994 which has an estimated age between 7 - 16 Myr \citep{1983ApJ...264..470H,1996AA...306..125C,2008MNRAS.386.1380K,2010AA...517A..50G} and contains a few YSG and RSG \citep{1979ApJS...39..389H,1982ApJ...253..580F}. HD 37974 appears to be a member of star cluster NGC 2050 (LH 93) which contains no early O-type stars and several mid and late O-type and early B-type stars \citep{1994ApJS...91..583H}.  HD 38489 is close to cluster LH 104 with an estimated age of 3-5 Myrs \citep{1985A&A...150..151T}.  So even though HD 37974 and HD 23489 are close in projection to early O-type stars they are also closely associated with more evolved populations. Figure \ref{fig:classicBesg} and Table \ref{tab:high_lum} show the results of the nearest neighbor analysis agree well with the spatial distribution maps.  The highest luminosity B[e]sg have a distribution that is most similar to post-main sequence early B-type supergiants in the same luminosity range like the classic and candidate LBVs. 

\vspace{2mm}

\begin{figure}[ht]
\epsscale{.55}
\plotone{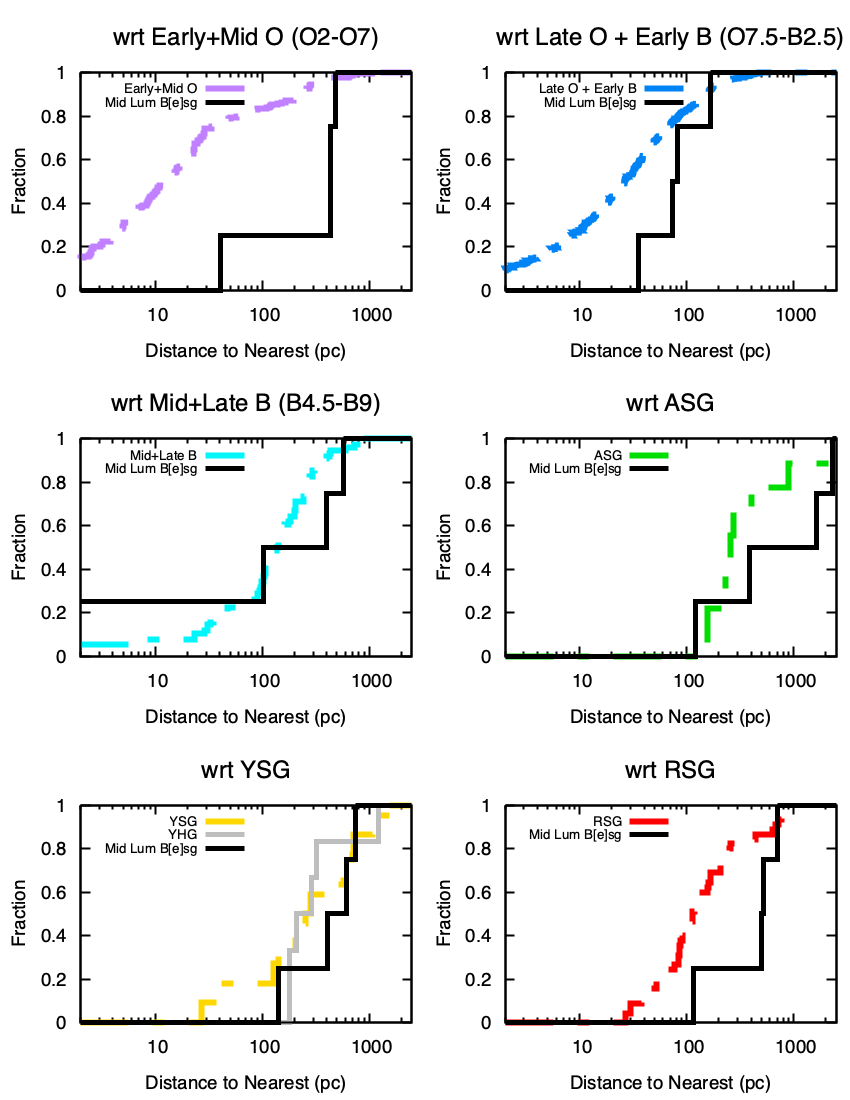}
\caption{Cumulative distribution plots of the nearest neighbors for mid-luminosity B[e]sg (black) plotted relative to the  mid luminosity $-8.0 > M_{Bol} \geq -9.7$ reference samples (colored dotted line). The YHG sample is also plotted relative to the YSG as a gray line in that panel.}
\label{fig:midlumBesg}
\end{figure}
\vspace{2mm}

The mid luminosity B[e]sg ($-9.7 \leq M_{Bol} < -8$) include a wider range of spectral types (B0 to A0).  Spatial plots show that three of the four mid-luminosity B[e]sg are well separated from any early O-type stars (Figure \ref{fig:LMCMap_midlum_zoom}). The exception, HD 269217, is on the edge of stellar association LH 35 which has an estimated age less than 10 Myr \citep{1985AA...152..427C,1996ApJS..102...57B}.  LHA 120-S-35 is associated with cluster KMHK 915 which has an estimated age of 10-30 Myr \citep{2005AA...442..597V,2018AA...612A.113T}. All four are in relative proximity to early B-type stars and LHA 120-S 35 and HDE 268835 are also close to A-supergiants, YSG, and RSG of similar luminosity.  Figure \ref{fig:midlumBesg} and Table \ref{tab:low_lum} show the results of the nearest neighbor analysis.  Their cumulative distributions show that the mid luminosity B[e]sg are dispersed similar to other post-main sequence samples. Although there is not a strong match the correlation is best with the late B-type supergiants.

\begin{figure}[t]
\epsscale{0.7}
\plotone{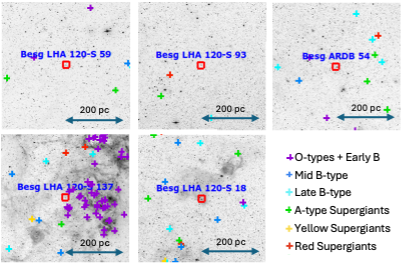}
\caption{The lowest luminosity B[e]sg plotted on H-alpha emission images of the LMC from MCELS \citep{1999IAUS..190...28S}.  The plots also include the reference samples in the matching luminosity range (Table \ref{tab:refsamp}): O-types and Early B-Types (purple), Mid B-types (blue), Late B-types (cyan), A-type supergiants (green), yellow supergiants (gold), and red supergiants (red)..}
\label{fig:LowestLumBesgMap}
\end{figure}

\begin{deluxetable*}{llrrrrrrccl} [t]
\tabletypesize{\scriptsize}
\tablewidth{0pt} 
\tablecaption{Distance to Nearest Neighbors for Lowest Luminosity B[e]sg\label{tab:besg_lo_d1}}
\tablehead{
&&\multicolumn{3}{c}{Mid B-type\tablenotemark{a}}
&\multicolumn{3}{c}{Late B-types\tablenotemark{a}}
&\\
&&
\colhead{D1}&\colhead{D2}&\colhead{D3}&
\colhead{D1}&\colhead{D2}&\colhead{D3}&
\multicolumn{2}{c}{Nearest\tablenotemark{b}}&\colhead{In Line of Sight\tablenotemark{c}}\\
\colhead{Star}&\colhead{Sp Type}&
\colhead{(pc)}&\colhead{(pc)}&\colhead{(pc)}&
\colhead{(pc)}&\colhead{(pc)}&\colhead{(pc)}&
\colhead{SpType}&\colhead{D(pc)}&
\colhead{Name}
}
\startdata 
ARDB 54&A0-II&134&244&277&153&156&192&B1&94&KMHK 264\\
LHA 120-S 93&A0:I:&257&346&403&184&285&317&B8&184&-\\
LHA 120-S 59&B0.5IIIe&117&188&197&140&176&258&B1.5&25&-\\
LHA 120-S 137&BIaePCyg&210&382&481&227&284&310&B1&111&-\\
\enddata
\tablenotetext{a}{Reference samples only include stars in the same luminosity range $M_{Bol} \geq -7$. D1, D2, and D3 are distances to the first, second, and third nearest stars in the reference sample in parsecs.}
\tablenotetext{b}{Spectral type and distance in parsecs to the nearest star in the same luminosity range.\\}
\tablenotetext{c}{Determined using \citet{2008MNRAS.389..678B}.}
\end{deluxetable*}

\begin{figure}[t]
\epsscale{0.62}
\plotone{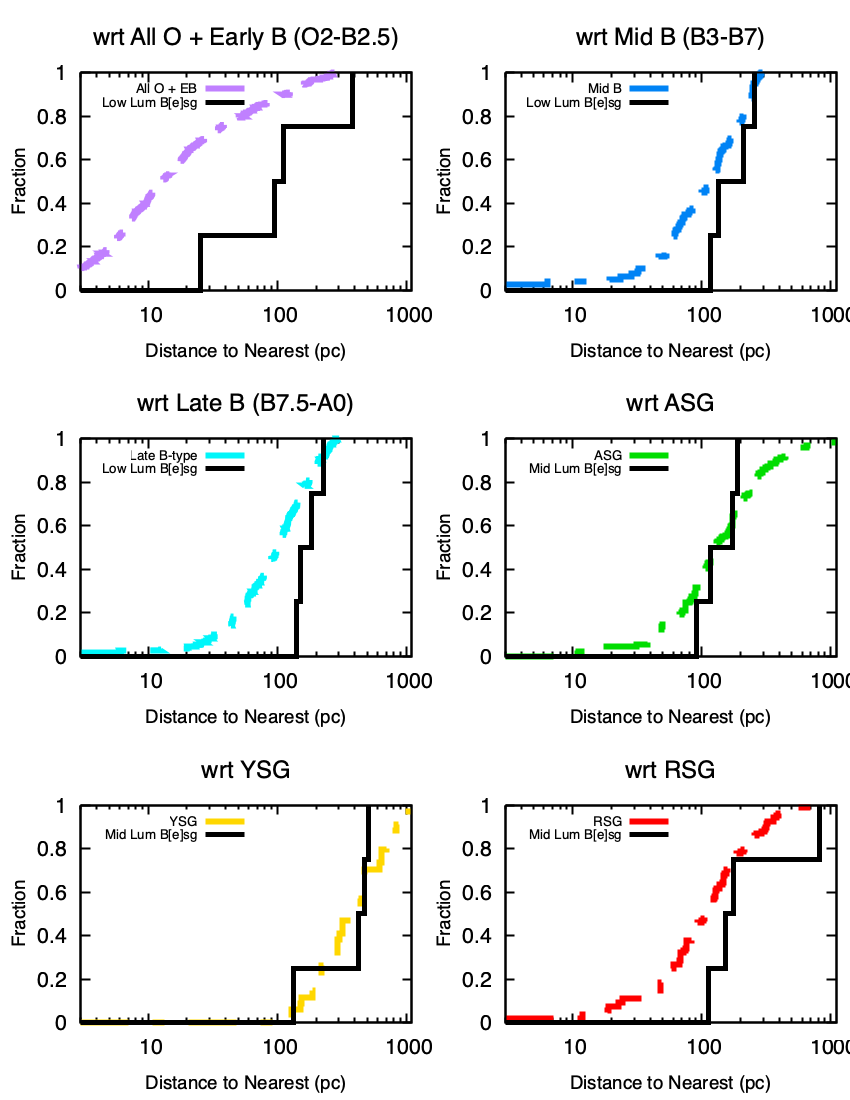}
\caption{Cumulative distribution plots of the nearest neighbors for low-luminosity B[e]sg (black) plotted relative to different lowest luminosity $-7.0 \leq M_{Bol} < -5.0$ reference samples (colored dotted line).}
\label{fig:latelumBesg}
\end{figure}

The lowest luminosity B[e]sg ($-7 \leq M_{Bol} > -5$) range in spectral type from mid B-type to A0. Note that we did not select for stars in this luminosity range for our HRD. Therefore, the reference samples are incomplete especially for the A-type and yellow and red supergiants at these luminosities. The spatial map comparing the distribution of lowest luminosity B[e]sg shows that they are mostly isolated with the exception of LHA 12-S-137 (Figure \ref{fig:LowestLumBesgMap}), which is on the edge of association LH 108, and close to late O-type and early B-type main sequence stars. ARDB 54 is associated with star cluster KMHK 264 which has an estimated age between 9-32 Myr \citep{2003AJ....126.1836H,2010AA...517A..50G,2012ApJ...751..122P}.  The nearest neighbor analysis is shown in Table \ref{tab:besg_lo_d1} and Figure \ref{fig:latelumBesg}.  The nearest star for each of the lowest luminosity B[e]sg are B-type stars close in spectral type to the target itself. The distribution of the lowest luminosity B[e]sg relative to each reference sample best matches the A-supergiants and YSG.

\subsection{Wolf-Rayet Stars}\label{analysis-wr}
Wolf-Rayet stars are found predominantly near other luminous stars and in the line of sight with young clusters and H II regions (Figures \ref{fig:wr-wo_zoom}, \ref{fig:LMCMap_WR-slash_zoom},  \ref{fig:wr_low_lum_Map}, and \ref{fig:wr_hi_lum_Map}).  This is consistent with the theory of WR as O-type stars which have been stripped of their hydrogen envelope \citep{1996LIACo..33..655C}.  \citet{2005A&A...429..581M} predict that  at the metalicity of the LMC,  the minimum mass for a WR progenitor in single star evolution is about 20 M$_{\odot}$ although, lower mass stars may become WR through binary interaction.  The same modeling also supports the possibility that some single star WR may become RSGs prior to entering their WR phase.  As a result, WR should be dispersed similar to other massive stars which have evolved past the main sequence including  possible post-RSG evolution. In the following discussion, we treat each subgroup of WR stars separately.  
 
\subsubsection{WN Stars}

The comparison between the WN samples and the reference populations is robust because there are several dozen stars in each set. The late WNs are considered to be the entry point to the WR sequence, and over time as they lose more of their hydrogen envelope, they evolve to the earlier type, hotter WNs \citep{1983ApJ...274..302C,1991A&A...249..443H}. The hotter early-type WNs may be more evolved or have been in the WR state longer than the later types with more hydrogen in their spectra. Thus the early and late WN stars may have different ages and different distributions. 

\vspace{2mm}

The  WNh stars are a subset of the WN classification which are  relatively hydrogen rich \citep{1973IAUS...49...15S, 1983ApJ...268..228C}.  They likely  have different origins depending on their luminosity.  The highest luminosity WNhs ($M_{Bol} < -9.7$) may be core hydrogen burning stars between $65 - 110 M_\odot$ \citep{1994A&A...290..819L,1995A&A...293..403C}.  While the lower luminosity ones ($M_{Bol} \geq -9.7$) are considered classical WN in a relatively early phase \citep{2006A&A...457.1015H,2014A&A...565A..27H} with CNO processed material in their atmospheres \citep{2008A&A...478..219M, 2009A&A...495..257M, 2022A&A...663A..36B}. In the following discussion they have been separated from the late and early WN types.  

\vspace{2mm}

\begin{figure}[t!]
\epsscale{.7}
\plotone{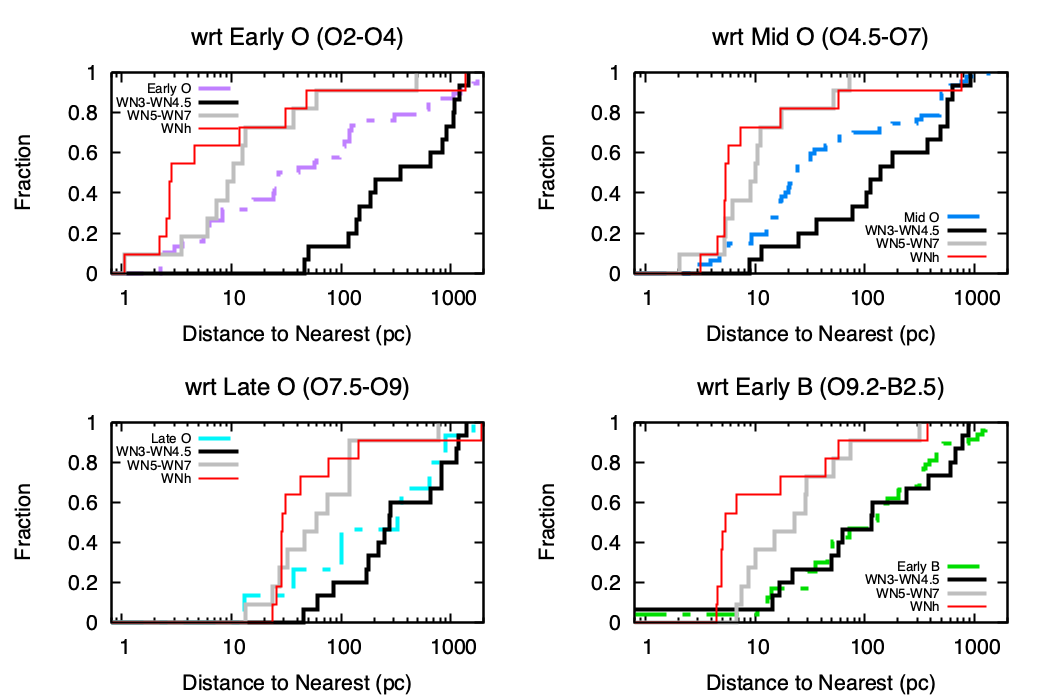}
\caption{Cumulative distribution plots of the nearest neighbors for the highest luminosity WN separated into groups by early types (WN3 - WN4, black) and later types (WN4.5 - WN7, gray).  The WNh (all later type) are plotted as a thin red line that overlaps the later type WN. Each sample is plotted relative to different high luminosity $M_{Bol} < -9.7$ reference samples (colored dotted line).}
\label{fig:wr_cumplot_hilum}
\end{figure}

We separated the high luminosity WNs by type into early (WN3 - WN4) and late (WN4.5 - WN7) with 15 and 11 stars, respectively. Figure  \ref{fig:wr_hi_lum_Map}   shows that the late WNs are highly concentrated in the 30 Doradus region while the early WNs are more spatially dispersed. The early WN's have a median separation of 370 pc relative to each other, which is comparable to the late O-type supergiants of the same luminosity (median separation of 329 pc, Table \ref{tab:refsamp}).  Figure \ref{fig:wr_cumplot_hilum} shows that the early WN's have a spatial distribution similar to evolved early B-type supergiants. The late WN's have a median separation of 262 pc relative to each other but do not closely match the spatial dispersions of any of the high luminosity O stars or early B supergiants. All eleven higher luminosity WNh are of later type (WNh4.5 -- WNh6) and all but one of them are found in and around 30 Doradus (Figure \ref{fig:wr_hi_lum_Map} and \ref{fig:LMCMap_WR-slash_zoom}). The higher luminosity WNh have a spatial distribution  similar to the later type WN, and like them do not match the distributions of the reference groups in the same luminosity range (Figure \ref{fig:wr_cumplot_hilum}). This reflects the concentraton of both groups in 30 Dor and suggests a possible close association and evolutionary link between them. 

\vspace{2mm}

\begin{figure}
\epsscale{.5}
\plotone{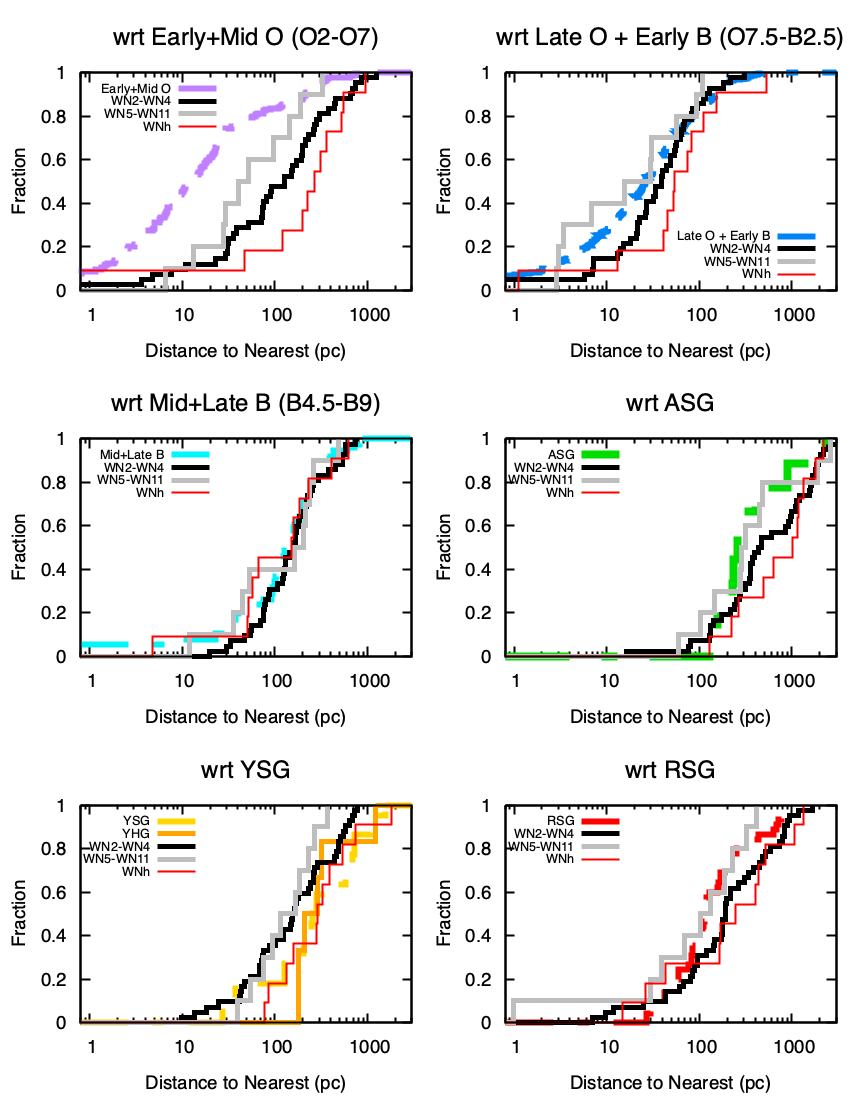}
% LMC_WRWN-early-late-wrtEach.bt-8to-9.7.png}
\caption{Cumulative distribution plots of the nearest neighbors for the less luminous WN separated into groups by early types (WN2 - WN4, black) and later types (WN5 - WN11, gray).  The distribution of less luminous WNh is plotted as a thin red line Each sample is plotted relative to different mid luminosity $-8.0 > M_{Bol} \geq -9.7$ reference samples (colored dotted line).}
\label{fig:wr_wn_cumplot_lolum}
\end{figure}

Figure \ref{fig:wr_wn_cumplot_lolum} shows the nearest neighbor distribution of the 52 less luminous WN split into earlier types (WN2-WN4) with 42 members  and later types (WN5-WN11) with 11 stars.  Both samples have similar spatial dispersions and compare favorably with the evolved supergiants from late B-type to the red supergiants.  Together all 52 less luminous WN have a median separation of 135 pc with respect to themselves, which is comparable to the median separation of similar luminosity mid + late B-types with respect to themselves (136 pc, Table \ref{tab:refsamp}). The eleven less luminous WNh are spread evenly across types WNh2 --- WNh7. Figure \ref{fig:wr_wn_cumplot_lolum} shows that like the less luminous WN, the spatial distribution of the less luminous WNh compare most favorably with evolved stars of similar luminosity including YSGs and RSGs.  
\vspace{2mm}

\cite{2018ApJ...863..181N} suggested that WN3/O3 stars represent a transitional state between early O-type stars with hydrogen envelopes and full-fledged WN. All the known WR stars of this sub-type are found in the LMC. Figure \ref{fig:wr_low_lum_Map} shows that WN3/O3 are usually found in proximity to the less luminous WN and spread over the whole area of the LMC.  This supports the finding by \cite{2018ApJ...863..181N} that WN and WN3/O3 are closely associated and could be drawn from the same population.  

\subsubsection{WC Stars}
The WC Wolf-Rayet stars are considered to be more evolved than WN because their spectra show the products of helium fusion \citep{1943ApJ....98..500G,1991ApJ...368..538L}, while the WN stars show the products of the CNO process. There are two WC stars in our sample with luminosities that place them only slightly above the HD limit.  We therefore treat all 23 WC as a single population for comparison with with the mid-luminosity reference samples  $-8.0 > M_{Bol} \geq -9.7$ (Table \ref{tab:refsamp}).  The mean separation between the WC with respect to each other is 162 pc, more than the mean separation of the  mid + late B-types (136 pc) and RSG (113 pc) but less than the ASG (258 pc) and YSG (282 pc) in the same luminosity range.

\vspace{2mm}
\begin{figure}[b!]
\epsscale{0.5}
\plotone{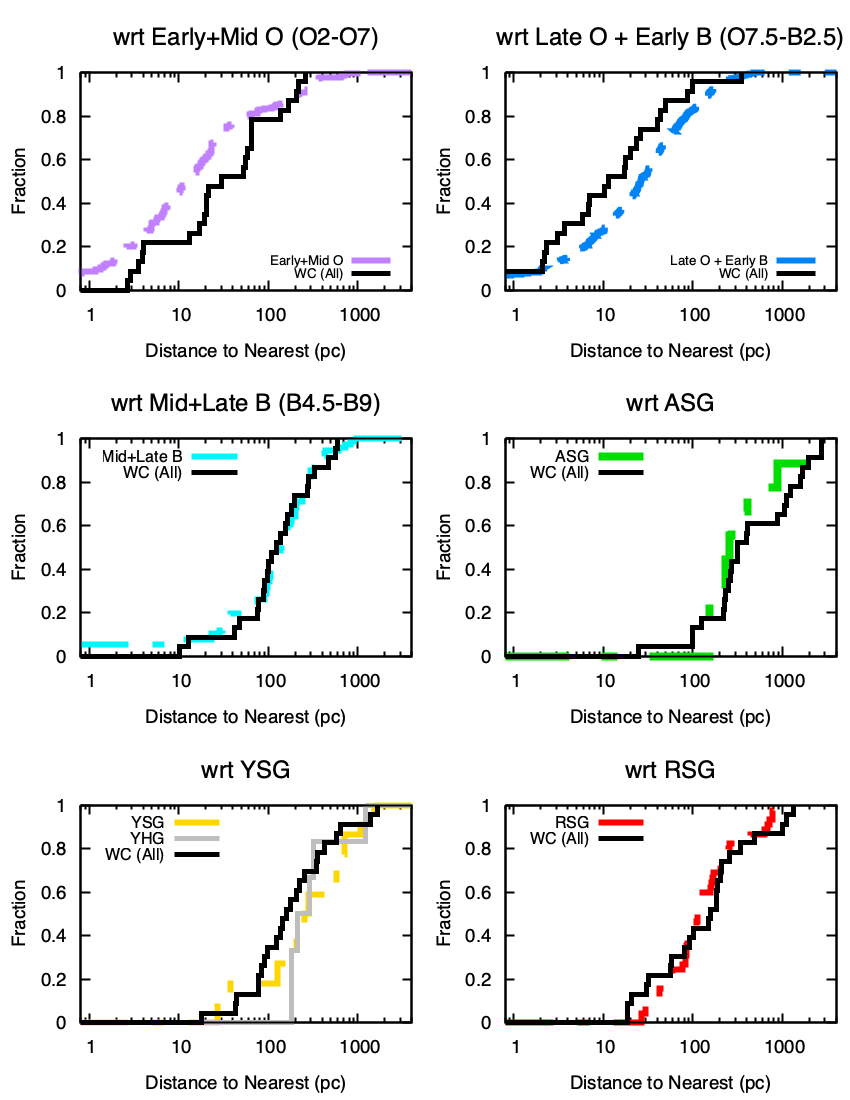}
\caption{Cumulative distribution plots of the nearest neighbors for all 23 WC (black) plotted relative to different mid luminosity $-8.0 > M_{Bol} \geq -9.7$ reference samples (colored dotted line).  The YHG sample is also plotted relative to the YSG as a gray line in that panel.}
\label{fig:wccumplotlolum}
\end{figure}

\begin{figure}
\epsscale{0.4}
\plotone{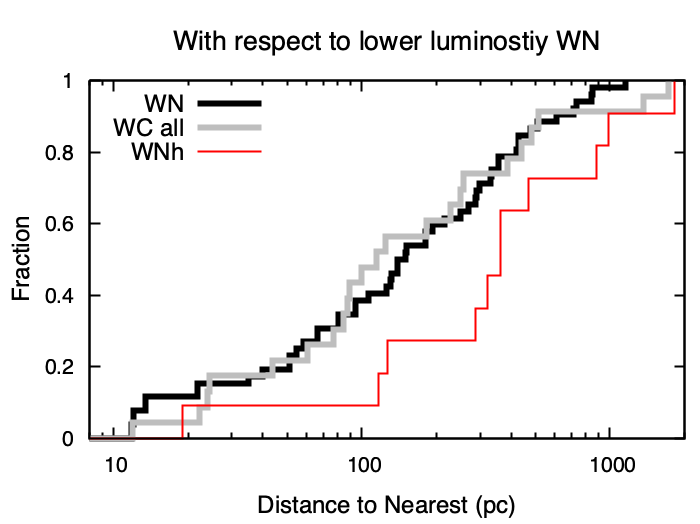}
\caption{Cumulative distribution plots of the nearest less luminosity $-8.0 > M_{Bol} \geq -9.7$ WN neighbors for the WN themselves (black), WNh (thin red line onr the right) and all 23 WC (gray).}
\label{fig:wr_wn_wc_cumplot_lolum}
\end{figure}

Figure \ref{fig:wccumplotlolum} compares the full sample of 23 WC to the reference populations between $-8.0 > M_{Bol} \geq -9.7$.   The distribution of the WC stars is interestingly consistent with the distribution of four evolved populations, the late B-type stars and the ASG, YSG, and RSG distributions.  A comparison of the less luminous WNs and WCs directly with each other shows that they have very similar spatial distributions (Figure \ref{fig:wr_wn_wc_cumplot_lolum}). However, the less luminous WNh, associated with the same reference populations as the WNs, are not  well-matched to the WNs. Probably because a significant fraction of the WNs and WCs are clustered together in the area around 30 Doradus, while most of the WNh are more dispersed (Figure \ref{fig:wr_low_lum_Map}). It is unclear how much of the match between the WC and WN spatial distributions, for example, is influenced by this clustering, which is a potential confusion factor and is most apparent in the WR samples. The WCs could be a post WN stage with one or both phases being relatively short lived so that the older phase does not have time to disperse significantly relative to the younger. Recall from Figure \ref{fig:wr_wn_cumplot_lolum} that the less luminous WNs are not as well matched to the ASG and RSG spatial distributions as the WC, supporting the expectation that WC are a later phase than WN.  

\begin{figure}[t]
\plotone{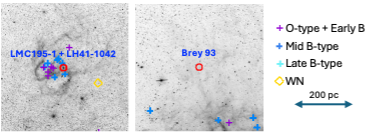}
\caption{The WO plotted with red circles on H-alpha emission images of the LMC from MCELS \citep{1999IAUS..190...28S}.  The plots also include the reference samples in the $-9.7 \leq M_{Bol} < -8$ luminosity range (Table \ref{tab:refsamp}): Early + Mid O-types (purple), Late O-types + Early B-types (blue), and Mid + Late B-types (cyan), and lower luminosity WN (gold diamonds).  There are no A-type supergiants, yellow supergiants, red supergiants or WC stars within the bounds of the area plotted around the WO stars.}
\label{fig:wr-wo_zoom}
\end{figure}

\subsubsection{WO Stars}
The three LMC WO ($-8.3 < M_{Bol} < -8.7$) are best compared with the mid-luminosity reference samples (Table \ref{tab:refsamp}). Unfortunately, there are too few WO for a meaningful nearest neighbor analysis. Figure \ref{fig:wr-wo_zoom} shows mid-luminosity reference stars and WR within 200 pc of each of the LMC WO stars.  [MNM2014]LMC195-1 and [L72]LH41-1042 are extremely close to each other and nearby several mid B-type stars of similar luminosity, the high luminosity LBV S Doradus and cluster LH 41.  Brey 93 is to the northeast of the Tarantula Nebula, isolated from all other stars in the same luminosity range.   \citet{2022ApJ...931..157A} suggests that WO stars are evolved WC stars reaching the end of a continuous sequence of evolution from late WC to early WO. However, there are no WC within 200 pc of the LMC WO stars.  The nearest neighbor analysis of WC show that they have a spatial distributions similar to the later-type evolved supergiants, while [MNM2014]LMC195-1 and [L72]LH41-1042 are associated with several evolved B-type stars but there are no nearby later-type supergiants. This could be a chance supperposition. Brey 93 is isolated which could support a more advanced state but it is difficult to tell since there only three WO in the LMC.

\begin{figure}
\plotone{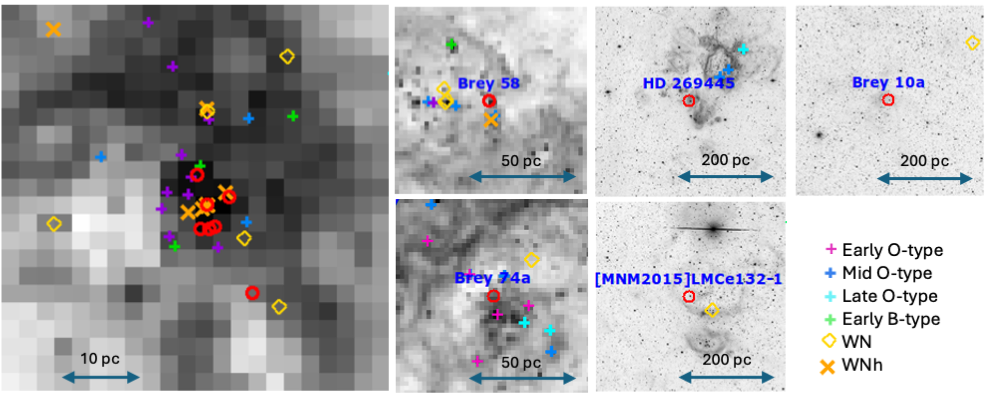}
\caption{The Of/WN slash stars plotted with red circles on H-alpha emission images of the LMC from MCELS \citep{1999IAUS..190...28S}.  The WN and WNh stars are plotted as gold diamonds and orange X's respectively.  The plots also include the reference samples in the matching luminosity range ($M_{Bol} < -9.7$): Early O-types (purple), Mid O-types (blue), Late O-types (cyan), and Early B-types (green).  Note the different spatial scales in each section.  The image on the far left is a zoom in on the 30 Doradus star forming region.}
\label{fig:LMCMap_WR-slash_zoom}
\end{figure}

\subsubsection{``Slash" Stars}
The Of/WN ``slash" stars have a composite spectral classification with an early-type O star and a late WN.  Figure \ref{fig:LMCMap_WR-slash_zoom} shows that they are associated with H-alpha emission and in proximity to WN's and other high luminosity stars.  The slash stars are the least dispersed of the WR samples with 8 out of 13 tightly grouped in the heart of 30 Doradus within a projected area fewer than 10 pc across.  Normally a small sample may have a large median separation because they would be widely scattered across the galaxy. Due to their concentration in 30 Dor,  their median separation relative to each other is only 4 pc, significantly less than 41 pc for early O-type supergiants and 24 pc for mid O-type supergiants in the same luminosity range (Table \ref{tab:refsamp}).

\vspace{2mm}

All but two of the slash stars (Brey 10a and HD 269445) are within 50 pc of higher luminosity WNs and WNhs, and except for [MNM2015]LMCe132-1, the nearest WN is a later type (WN4-WN11). The close association with later type WN is consistent with the sequence proposed by \citet{2011MNRAS.416.1311C} which identify the slash stars as a transitional stage between the main sequence and late-WN for very massive stars ($180-120 M_\odot$). The concentration of the high luminosity WNh and slash stars together in the center of 30 Doradus strongly associates these classes with each other.  Their highly concentrated distribution of may indicate youth compared to the main sequence stars with the same luminosity.   Their high density in 30 Doradus suggests they could be influenced by the exceptional nature of that star forming region.  Or there could be a selection bias causing the sample of Of/WN and higher luminosity WNh in the LMC to be incomplete if these stars are primarily found at the unresolved hearts of star forming regions.

\section{Summary and Discussion}

In this paper we have examined the spatial distributions of evolved massive stars, the LBVs and B[e]sgs and the different subgroups of Wolf-Rayet stars. The B[e]sg, LBVs, and Wolf-Rayet phenomena are all observed across a wide range of luminosities on the HR Diagram implying that stars with different initial masses may exhibit the associated features and characteristics of the different groups at different stages in their post-main sequence evolution. 

\vspace{2mm}

The classic LBVs, the LBV candidates,  the most luminous B[e]sgs and the high luminosity WN stars, have luminosities that place them above the H-D limit at M$_{Bol}$ $\approx$ -9.7 (Log L/L$_{\odot}$ $\approx$ 5.8), see Figures \ref{fig:hrd_lbv}--\ref{fig:hrd_wr},  with likely initial masses above 40 M$_{\odot}$. Their spatial distributions in Figure \ref{fig:LMCMap_hilum_zoom}, show that the three classic LBVs are observed projected towards H II regions and young clusters. Candidate LBVs are also observed near young clusters and emission regions except for S61 which  appears relatively isolated.  The high luminosity B[e]sg are  similarly associated with regions of young stars. 
   
\vspace{2mm}

Although the nearest neighbor analysis for the classic LBVs is limited to only three stars, they and the CLBVs have spatial dispersions that are more  consistent with the distribution of the luminous late O-type stars and early B-type supergiants. The B[e]sgs are also most closely associated with the early B-type supergiants, consistent with their early B-type spectral types. Their positions on the HRD (Figures \ref{fig:hrd_lbv} and \ref{fig:hrd_besg}) support a post main sequence state for the classic LBVs and the high luminosity B[e]sg. The well known nitrogen and helium-rich ejecta of the LBVs is also strong support for their  possible  post-CNO cycle state. They  may already be in He-burning. In contrast, the nearest neighbor analysis of the the highest luminosity WN stars  show mixed results.

\vspace{2mm}

The spatial distribution and nearest neighbor analysis yields some interesting results for the LBVs, and B[e]sg below the H-D limit in the luminosity range from M$_{Bol}$ -8.0 to -9.7 mag, corresponding to an initial mass range from $\approx$ 25 -- 40 M$_{\odot}$.  Based on the positions of the less luminous LBVs and B[e]sg on the HR Diagram we expect them to be post-main sequence stars older than their more luminous, more massive counterparts and likely more dispersed than O and early B-type stars of similar mass and luminosity.  The space distribution of both groups (Figure \ref{fig:LMCMap_midlum_zoom}) confirms that they are more dispersed and associated with more evolved supergiants. The LBV R71 and the B[e]sg LHA 120 S-12 appear to be the most isolated. The nearest neighbor analysis of the less luminous LBVs is consistent with their appearance in the spatial distribution map and with the distributions of evolved A-type supergiants and yellow supergiants (Figures \ref{fig:lesslumLBV} and \ref{fig:midlumBesg}). Interestingly, their distribution bears a resemblance to the distribution of post-RSGs yellow hypergiants in the YSG panel of Figure \ref{fig:lesslumLBV}. The cumulative distribution of the less luminous B[e]sg, is also similar to the more evolved late B supergiants, A-type and yellow supergiants, although the correlation is not as strong as for the less luminous LBVs. 

\vspace{2mm}

Evolutionary tracks, with and without rotation, for stars in this luminosity and mass range \citep{2021A&A...652A.137E} show that they may cross the region of the evolved yellow and blue supergiants more than once. They may spend as much time on a post-RSG blue loop as on their first crossing of the HR Diagram. Both evolutionary states occupy the same temperature and luminosity locus, and although this analysis is not adequate to distinguish them, the less luminous LBVs may be post-RSGs. 

\vspace{2mm}

The lower luminosity B[e]sg are the lowest luminosity group included in this study.  Their luminosities correspond to initial masses of about 10 -- 12 M$_{\odot}$.  Their spatial distribution show that they are mostly isolated and their nearest neighbor distributions are, not surprisingly, most similar to the A-type and yellow supergiants of the same luminosity. At these masses and luminosities, the stars are not expected to have high line-driven mass loss like luminous O and B-type stars or experience high mass loss episodes like the LBVs. Thus their dusty envelopes and evidence for mass loss may be the consequence of rapid rotation or binarity.

\vspace{2mm} 

As discussed earlier, the evolutionary state of the B[e]sg is open for debate.   Their wide range of luminosities suggests the cause of their common observed properties is not a strong function of their mass or luminosity.  Their dusty equatorial regions may be due to rotation, binarity or both. Clearly the most luminous B[e]sg in our sample have not been RSGs. The mid-luminosity B[e]sg however are dispersed similarly to the evolved late B-type, A-type, and yellow supergiants. They appear to be more dispersed and more isolated than the less luminous LBVs with similar luminosities. The lowest luminosity group is obviously an older subset most consistent with the A-type and yellow supergiants. Thus, based on their in-common characteristics, stars from a wide range of initial masses can display the B[e]sg phenomena along more than one evolutionary path.

\vspace{2mm}

\begin{figure}[t]
\epsscale{0.40}
\plotone{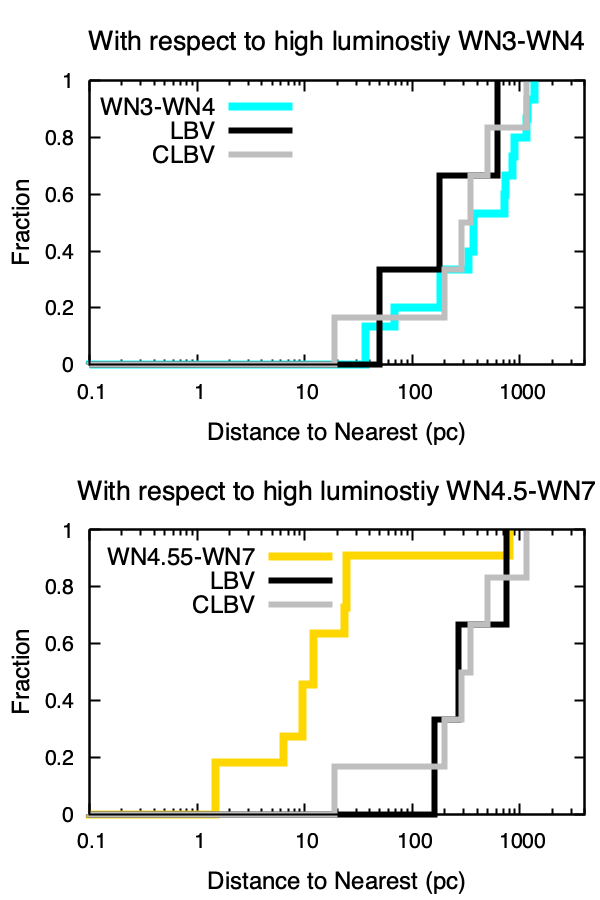}
\caption{Cumulative distribution plots of the nearest neighbors for the classical LBV (black) and candidate LBV (gray) plotted relative to early (WN3-WN4) and late (WN4.5-WN7) higher luminosity ($M_{Bol} < -9.7$) WN stars (colored line).}
\label{fig:LBV_WN}
\end{figure}

The standard explanation for the WR stars is that they are OB stars which have been stripped of their hydrogen envelopes either through mass loss or binary interaction.  The high luminosity earlier type WN (WN3-WN4) have a distribution similar to evolved late O-type supergiants while the later type WNs (WN5-WN7) and WNh stars are less dispersed than the early types and are not associated with any of the reference populations (Figure \ref{fig:wr_cumplot_hilum}). This appears  to call into question the expected relationship between later type WNs and LBVs. For example, two of the classic LBVs and two candidates are classified as late WNs in their quiescent states. Consequently, it was suggested that the high luminosity LBVs would eventually become late WNs.  The WN spectral characteristics in the LBVs are likely due to the decrease of their H-rich envelopes through LBV mass loss episodes.  The nearest neighbor analyses in Figure \ref{fig:LBV_WN} clearly show a similar spatial distribution between classic LBV and the early-type WNs. The late WN and WNh distributions, however, confirm what we saw in Figure \ref{fig:wr_cumplot_hilum} with the strong qualification that it reflects their high concentration in the 30 Dor region which can bias the conclusions from the cumulative distribution plots as discussed in \S {3}. It is not surprising that the LBVs and high mass losing early WNs are correlated but it is therefore premature to conclude that the high luminosity LBVs do not first pass through the late-WN stage, even briefly.

\vspace{2mm}

The late- and early-type less luminous WNs, below the H-D limit, show similar spatial distributions. Their spatial distribution map (Figure \ref{fig:wr_low_lum_Map}) shows a strong correlation with the H II regions and their nearest neighbor cumulative distributions are consistent with the late B-type supergiants, and like the less luminous  LBVs, also with more evolved A-type and yellow supergiants. The  early-type less luminous WNs have presumably shed most of their H-rich envelopes, however the standard mass loss prescription for B-type supergiants from line-driven winds is not high enough in this luminosity range, and even less so with clumping, to strip their outer envelopes. In this luminosity range, mass exchange in binaries is a more plausible path from evolved B-supergiant to WN, perhaps similar to the B[e]sg. Alternatively, it is often suggested that the less luminous LBVs will eventually become WNs via their high mass loss episodes.  We note that three of the late WNs (Table \ref{tab:wr_wn}) are classified W10/WN11 and may be found someday to be LBVs or candidates. The correlation of both groups with the more evolved supergiants as well as possible post-RSG evolution supports this outcome. We do not have sufficient data to distinguish among these possibilities, but the data at hand favor a post-LBV or post-RSG origin for the less luminous WNs.

\vspace{2mm}

The WC stars have a similar spatial distribution to stars which could have evolved through the yellow and red supergiant phases (Figure \ref{fig:wccumplotlolum}), and Figure \ref{fig:wr_wn_wc_cumplot_lolum} shows that the WC and less luminous WN stars have been drawn from the same population. Evolutionary tracks suggest the few WC above the HD Limit could have evolved to their present point from less luminous post-RSG progenitors. Combined with what is already established by other means, including abundance analysis, the information gleaned from the WN and the WC distributions favor the standard scenarios in which the less luminous WN and the WC stars are drawn from the same populations with the WC stars representing a more evolved state showing the products of He-burning.

\vspace{2mm}

Our results for the other WRs show differences and possible associations between the sub-classes of stars exhibiting the Wolf-Rayet phenomenon.  “Slash” stars and high luminosity WNh are most associated with very recent high mass star formation, almost all of them in a projected area 10 pc across in the heart of 30 Doradus. Both the Of/WN slash stars and high luminosity WNh are highly concentrated in the core of 30 Doradus. This supports the suggestion by \cite{1998ApJ...493..180M} that these are relatively unevolved stars and the evolutionary link between slash stars and WN proposed by \cite{2011MNRAS.416.1311C}.  However, as noted previously, the core of 30 Doradus is heterogeneous with a spread of different aged populations. Since the Of/WN slash stars, the high luminosity WNh and late WNs are significantly concentrated in one unique star forming region there is a potential for bias when comparing them with populations spread over different size scales in the LMC. The lower luminosity WN ($-8.0 > M_{bol} \geq -9.7$), WC and WN3/O3 show indications that they are drawn from the same population reinforcing the suspected evolutionary links between them (Figures \ref{fig:wr_wn_cumplot_lolum} and \ref{fig:wr_wn_wc_cumplot_lolum}). The WO sample is too small to draw clear conclusions.  

\subsection{Transverse Velocities of LBVs and B[e]sg}\label{vreltrans}

In a recent paper, \citet{2024ApJ...976..125D} argue that a subgroup of LBVs, candidate LBVs, and B[e]sg in the LMC are accelerated and the products of binary mass transfer. They used the GAIA DR3 proper motion catalog to measure the transverse velocities of LBVs and B[e]sg  and then compared their motions with a sample of OBe stars in the SMC. Here we show that as a class most LBVs and B[e]sgs are not associated with runaways. 

\vspace{2mm}

Deman and Oey adopt the list of LBVs, candidates, and B[e]sgs from \citet{2019AA...626A.126A} with some additions and corrections. As mentioned in \S \ref{sec:lbv} that list includes two non-LBVs and two $\alpha$ Cyg variables\footnote{\label{foot:deman} HD 268939 (R 74), HD 269050 (R 78), HD 269128 (R 81),and HD 269445 (R 99). Interestingly, LHA 120-S 18 in their list may be a case of mistaken identity. A star labeled S 18 is included in \citet{2001AA...366..508V} as an LBV candidate, but all of the references are to LHA 115-S 18 in the SMC, a well known B[e]sg \citep{1989A&A...220..206Z}. LHA 120-S 18 in the LMC is an emission line star but there are few published references to its spectrum and none to variability.} They also adopt the dust-based classification from \citet{2019AA...626A.126A} which is based on the presence of or lack of mid-IR dusty emission. This classification scheme disregards the astrophysically important information from their positions on the HR Diagram; their temperatures, luminosities, possible evolutionary state, and associated stellar population. 

\vspace{2mm}  

Furthermore, many of the LBVs with dusty emission are in H II regions which also have a dusty signature. For example, R143 has a dusty nebula with an estimated dust mass of 0.055 $M_{\odot}$ or a total mass, dust plus gas, of approximately 5.5 $M_{\odot}$ \citep{2019AA...626A.126A}; high for an LBV. Assuming a nominal mass loss rate of $\approx$ 5 $\times$ 10$^{-5}$ M$_{\odot}$ per year during its high loss stage and it is in this state about half the time, it would take 10$^{5}$ years for R143 to shed that much mass. \citet{2019AA...626A.126A} acknowledge that R143 is not sufficiently luminous to ionize the nebula.  Thus the  dusty excess in its SED is probably associated with star formation in the H II region. Two of the CLBVs and the LBV R71 with dusty emission  are not identified with H II regions. Alternatively, the dusty excess may be a residual from a prior RSG state, a possibility for the less luminous LBVs.

\vspace{2mm} 

To calculate the relative transverse velocities for comparison, Deman and Oey selected a reference field population with Gaia magnitude G $<$ 18 within three 
arc minutes of the targets ($\sim$ 5600 pc$^{2}$ at the distance of the LMC). The number of field stars compared to each target is typically more than 100 and, for many cases, several hundred stars. The brightness criteria (Gaia magnitude G $<$ 18) ensures this will be a very mixed population with a wide range in luminosities including stars much fainter and less massive than the LBVs and B[e]sg. 

\vspace{2mm}

Throughout this work we have emphasized the importance of comparing these evolved stars with the population of stars that share their locus in the HRD. As 
the stars disperse with age we should expect the stars they are born with to be equally dispersed. Their spatial distributions and cumulative distributions in the nearest neighbor analysis confirm this. 
Consequently, we need to compare the motions of the target stars  with similar luminosity stars. To properly analyze their motions we need to know more about the reference populations; the spectral types and luminosities in addition to the transverse velocities and errors, and the velocity spread for the field stars.  Furthermore, to compare the target's velocities with models for competing ejection mechanisms requires measuring their velocities relative to their point of origin. The LMC reference field established by Deman and Oey does not appear to be consistent with that requirement.  

\vspace{2mm}

Instead, Deman and Oey compare the LMC stars with a survey of OBe stars in the SMC \citep{2024ApJ...966..243P} which is not only a different, smaller galaxy, but a significantly different stellar population with only one confirmed LBV, R40. The full OBe sample is heavily biased towards lower mass, less luminous stars. \citet{2024ApJ...966..243P} acknowledge that their criteria are also biased against high mass walkaway stars and note that the more massive stars in their runaway SMC OBe sample are moving faster than the average computed for the full sample. Nevertheless, Deman and Oey conclude that the LBVs and candidates in their class 1 with dusty emission and the B[e]sg have velocity distributions similar to the OBe runaways in the SMC. They comment that their second group, class 2, with no dust, does not show any evidence for acceleration. Class 2 includes the LBV HD269582, which has a high velocity (Figure 20), and two LBV candidates, but the other four members are neither LBVs nor candidates. See the footnote \ref{foot:deman} at the beginning of this section. Class 2 is not a realistic subset.

\vspace{2mm}
\begin{figure}[t]
\epsscale{0.45}
\plotone{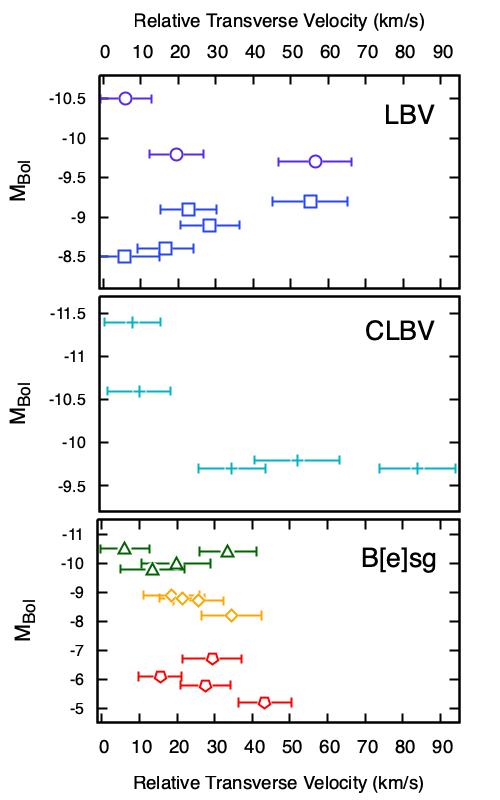}
\caption{Luminosity (M$_{Bol}$) as a function of relative transverse velocity \citep{2024ApJ...976..125D} with one sigma error bars for: LBVs (top panel), Candidate LBVs (middle panel) and B[e]sg (bottom panel). 
In the top panel: higher luminosity LBVs (purple circles, $M_{Bol} < -9.7$) and less luminous LBVs (blue squares, $M_{Bol} \geq -9.7$). In the bottom panel: high luminosity B[e]sg (green triangles, $M_{Bol} < -9.7$), mid-luminosity B[e]sg (orange diamonds, $-9.7 \leq M_{Bol} < -7.0$), and low luminosity B[e]sg (red pentagons, $M_{Bol} \geq -7.0$).
\label{fig:LBV_vrel}}
\end{figure}
\vspace{2mm}

For these reasons, we re-examine the Deman and Oey results using the LBVs, candidates, and B[e]sgs from Tables \ref{tab:lbv} and \ref{tab:besg} in this paper with the classification based on their positions on the HRD.  Figure \ref{fig:LBV_vrel} reveals a somewhat different, more nuanced picture of the velocity distributions of the LBVs, candidate LBVs,  and the B[e]sg. Their   luminosities are shown  relative to their transverse velocities from Table 3 in  \citet{2024ApJ...976..125D}. Most of the confirmed LBVs have transverse velocities less than 30 km s$^{-1}$. The two with higher velocities between 50 and 60 km s$^{-1}$ are the less luminous LBV R71 which we have previously noted is isolated compared to the stars of comparable luminosity and temperature, and the LBV HD 269582 which although not isolated has the largest separation from stars of comparable luminosity for the three classic LBVs. The candidates include two stars with high velocities; the very high velocity and apparently isolated S61 and S119 which is associated with a somewhat less luminous population mentioned in \S {3.1}. In contrast, the B[e]sg  show a relatively random distribution of velocities with luminosity, with a wide spread for each luminosity subgroup.

\vspace{2mm}

The velocity cutoff for candidate runaway stars suggested by \citet{2024ApJ...966..243P} in their SMC OBe sample is 24 km s$^{-1}$ . Adopting this velocity as a marker,  the average relative transverse velocities for the two high luminosity classic LBVs is 13 km s$^{-1}$. The four less luminous LBVs with lower masses and somewhat longer lifetimes have a slightly higher average at 18.4 km s$^{-1}$, although one has a transverse velocity slightly above the 24 km s$^{-1}$ cut-off.  Thus the majority of the confirmed LBVs are not candidate runaway stars.  The two most luminous candidate LBVs are likewise not runaway stars, although as a set, the candidates include three stars with velocities above the suggested cutoff for runaways.  The B[e]sg at all luminosities, have some stars above this nominal 24 km s$^{-1}$ with about half of the lowest luminosity set somewhat above this cutoff. These stars are older than the more luminous B[e]sg, and  we consider the B[e]sg as a group to be an older population than the LBVs, and therfore more dispersed. There is no evidence from their velocities to conclude that the B[e]sg are runaways.

\begin{figure}[b]
\epsscale{0.6}
\plotone{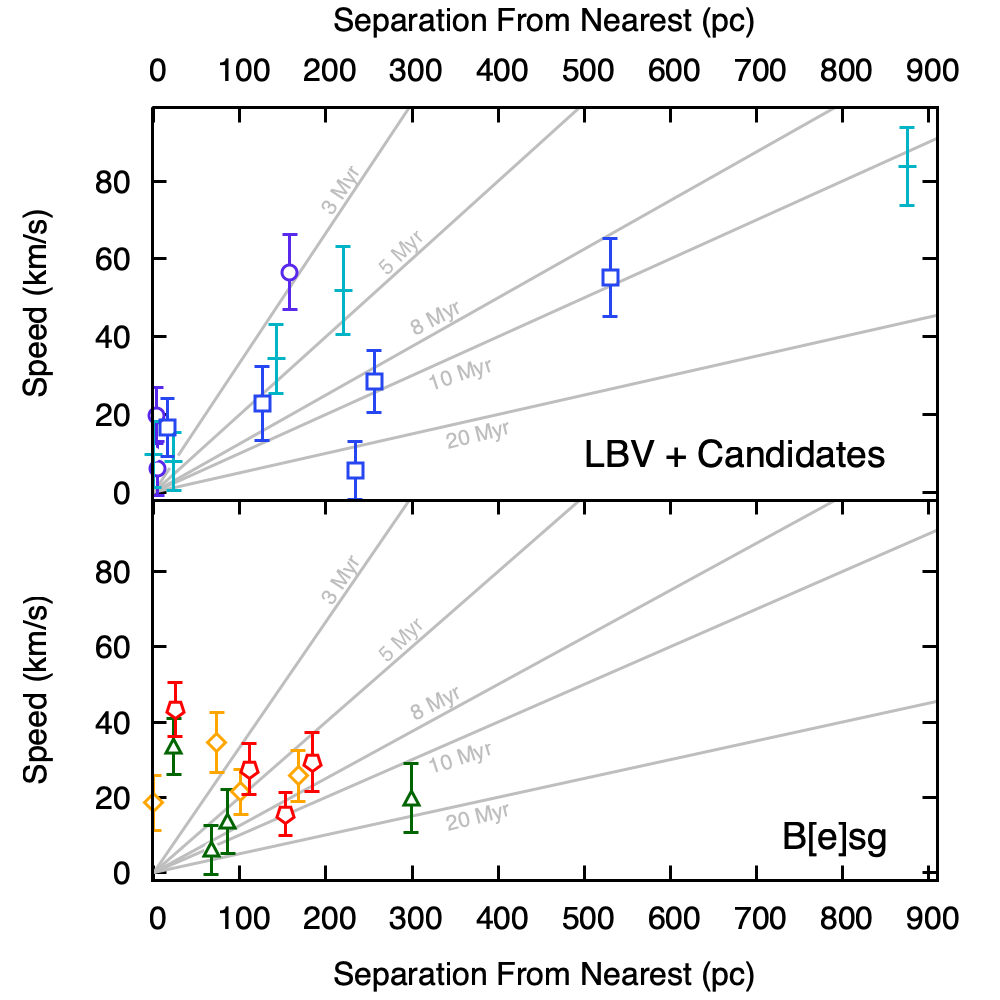}
\caption{The relative transverse speed with one sigma error bars computed by  \citet{2024ApJ...976..125D} as a function of separation from the nearest neighbor of similar luminosity and spectral type (Tables \ref{tab:high_lum} and \ref{tab:low_lum}).  The gray lines note the distance a star will move in 3, 5, 8, 10, and 20 Myr at a given speed. The colors for each class and luminosity grouping are the same as used in Figure \ref{fig:LBV_vrel}. In the top panel: higher luminosity LBV (purple circles, $M_{Bol} < -9.7$),less luminous LBVs (blue squares, $M_{Bol} \geq -9.7$), candidate LBV (cyan crosses). In the bottom panel: high luminosity B[e]sg (green triangles, $M_{Bol} < -9.7$), mid-luminosity B[e]sg (orange diamonds, $-9.7 \leq M_{Bol} < -7.0$), and low luminosity B[e]sg (red pentagons, $M_{Bol} \geq -7.0$).}
\label{fig:SepvSpeed}
\end{figure}
\vspace{2mm}

\begin{figure}[t!]
\epsscale{0.53}
\plotone{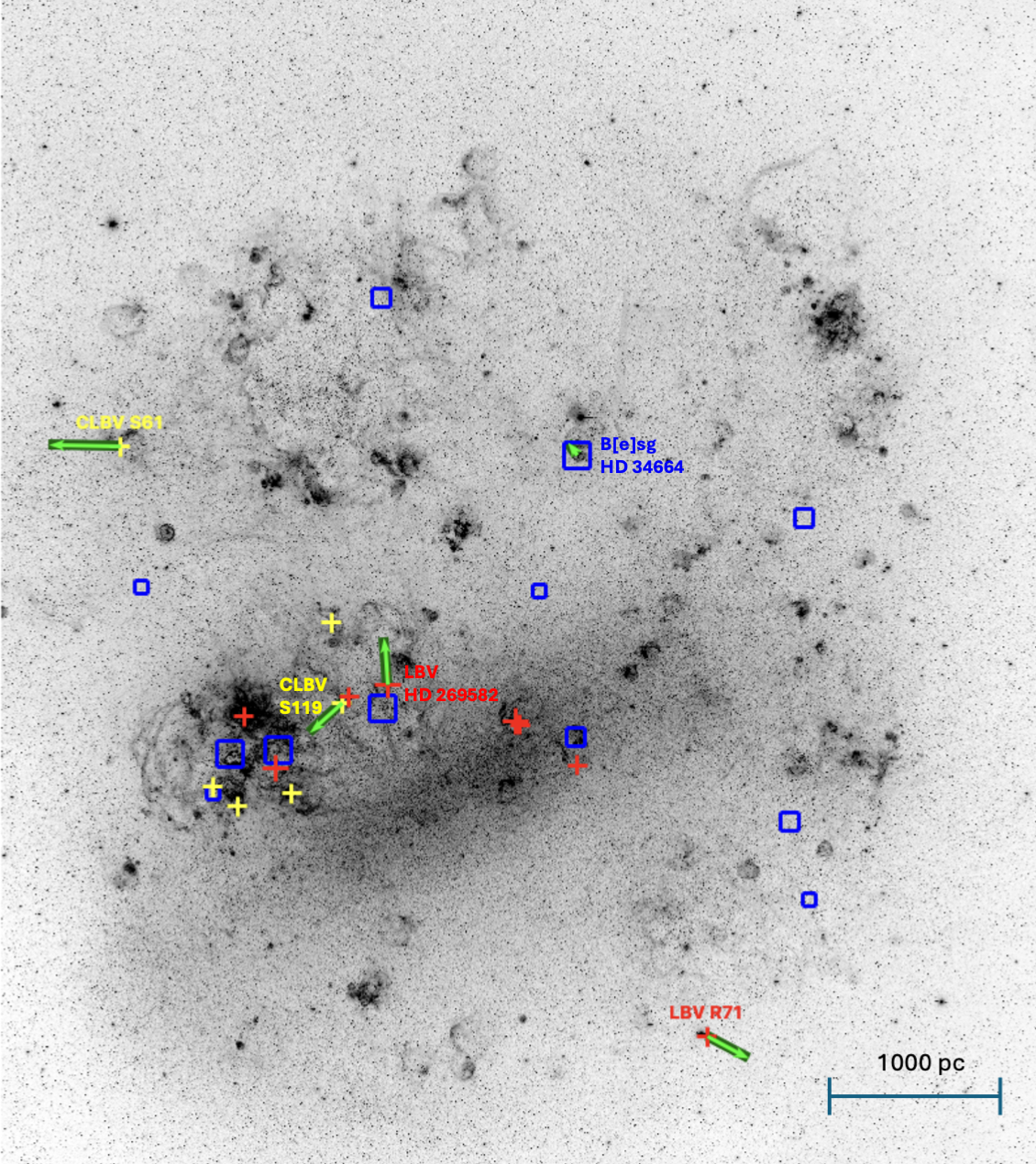}
\caption{An H-alpha emission image of the LMC from MCELS \citep{1999IAUS..190...28S} including LBVs (red crosses), candidate LBVs (CLBVs, smaller yellow crosses) and B[e]sg (blue boxes) in the LMC. The largest symbols for the LBV and B[e]sg are for those with $M_bol < -9.7$. Next largest symbols for those between $-8 > M_{bol} \geq 9.7$. And the smallest boxes for B[e]sg with $M_{bol} > -7$.  The green arrows show the direction and amount of transverse motion that the highest velocity stars and most separated B[e]sg (labeled) will experience over 5 Myr at the relative transverse velocity that \citet{2024ApJ...976..125D} has computed for them.}
\label{fig:LBV+BesgMap}
\end{figure}

As another check on the interpretation of the velocities, in  Figure \ref{fig:SepvSpeed} we  compare the relative velocities of each star with the separation from their nearest neighbor of similar luminosity and spectral type.  The nearest neighbor is {\em not} necessarily the point of origin, so this cannot be used to place precise limits on the ages of the individual stars.  However, Figure 21   reveals an important trend; younger populations need faster speeds to achieve larger separations.  That trend is present in the LBVs and candidate-LBVs, implying they are a younger population than the B[e]sg, which have slower speeds even at larger separations.  This is also supported by Figure \ref{fig:LBV+BesgMap} which shows that B[e]sg are widely spread across the LMC, as is expected for an older population.

LBV HD 269582 and candidate S 119 have relatively high velocities but could traverse the separations with their nearest neighbors in 3-5 Myr.  \citet{2021ApJS..252...21H} note the nebula around S 119 appears to be in a bubble blown by the star itself with a sharper, denser edge in the direction of the star's motion. While they could be runaways, their travel times imply they could be a product of single star evolution. There is no obvious contradiction which requires rejuvenation through mass transfer as advocated by \citet{2015MNRAS.447..598S}, \citet{2019MNRAS.489.4378S}, \citet{2022MNRAS.516.2142A}, and others. The high luminosity B[e]sg, HD 34664, has a greater separation from similar stars than the other B[e]sgs, and may have received some additional acceleration from binary interaction since binarity is a possible explanation for these stars' properties. 

\vspace{2mm} 

LBV R71, lying below the 8 Myr line in Figure \ref{fig:SepvSpeed}, is clearly the exception among LBVs.  It has a separation of 530 pc from its nearest neighbor of similar type which would take almost 10 Gyr to traverse at 55 $\pm$ 10 km s$^{-1}$. This would be hard to explain at its luminosity even as a post-RSG.  \citet{2017A&A...603A..75L} also noted that R71 is "rather unique in its isolation" with an apparent discrepancy between its probable age and time to travel from a group of OB stars (see Figure \ref{fig:LMCMap_midlum_early}) 450 pc to the north of R71. 

\vspace{2mm}

Candidate S61 is similar. Its high velocity (83.8 km s$^{-1}$) and isolation compared to the other candidates, suggests it may be a runaway.  While it is isolated from other high luminosity stars, there are three less luminous ($M_{Bol} > -9.7$) WN and six less luminous OB stars within 100 pc.  It is also associated with loose filamentary H II emission DEM L 308 / [KDS99] SGS 94 \citep{1976MmRAS..81...89D,1999AJ....118.2797K}. \citet{2021ApJS..252...21H} classified the nebula as a "large structure (i.e. super bubble)" about 85 pc across with an expansion age between 2-4 Myr embedded in an H I shell.  While there is an ionizing association of OB stars on its rim to the northwest, this structure may be a supernova remnant. The origin of S61's high velocity may be associated with this expanding nebula. 

\vspace{2mm}

Among the LBVs and candidates, R71 and S61 are the only two that may require additional acceleration.  Both dynamic ejection and the binary supernova scenario however, do not inevitably result in significant mass gain. In fact, abundance analysis of runaway OB stars in the solar neighborhood, including those likely to have been accelerated by the binary supernova scenario, show {\em no} evidence of processed material transferred onto their surfaces \citep{2006AJ....131.3047M,2017ApJ...842...32M,2023MNRAS.519..995L}. 

\section{Conclusions}\label{conclusions}

The following summarizes our conclusions.

\begin{itemize}
\item The majority of LBV/S Dor variables and candidates are not isolated from stars of similar luminosity and spectral type. The majority of confirmed LBVs are not runaways.

\item LBV R71 and candidate S61  have high velocities that require acceleration.

\item The nearest neighbor discussion of the B[e]sg is consistent with stars of the same spectral type and luminosity range although their spatial distribution and velocity spread suggests that they are an older population than the LBVs.

\item The early-type high luminosity WNs are correlated with the late O-type stars and with the most luminous LBVs. Conclusions about the late WNs and WNh stars are limited by their high concentration in the 30 Dor region.

\item The less luminous WNs, WNh stars,
	and WCs are most closely associated with more evolved stars in the 20 – 40 $M_\odot$ range including late-type B,  A-type supergiants and yellow hypergiants.

\item Based on the nearest neighbor analysis, a post-yellow hypergiant origin should be serioulsy considered for the less luminous LBVs, less luminous WNs and WCs.

\end{itemize}

\begin{acknowledgments}
 Martin's contribution to this work is supported by the Henry R. Barber Astronomy Observatory Endowment at the University of Illinois Springfield. We also thank Phil Massey for his feedback on this work and the anonymous referee for their constructive comments. 
\end{acknowledgments}

\appendix
\section{Updates to the Catalog of Luminous Stars in the LMC\label{hrdupdate}}

Twenty (25) stars from \citet{1999AJ....118.1684W}, 12 stars from \citet{2010AJ....139.1283W}, 6 stars from \citet{2014AA...570A..38B}, and 21 stars from \citet{2021A&A...648A..65C} with good spectral types and photometry have been added to the list of OB-type supergiants (Table \ref{tab:obstars}) and OB-type known and suspected binaries (Table \ref{tab:obabin}). When available, luminosity and effective temperatures from \citet{2014A&A...564A..40W} were adopted.  Otherwise physical parameters are derived from their spectral type according to the procedure in \citet{2023AJ....166..214M}.   

\vspace{2mm}

The intrinsic colors and bolometric corrections for O2 -- O3.5 types were revised according to \citet{2013A&A...558A.134D} and \citet{2017A&A...601A..79S} (Table \ref{tab:cal}).  Updated parameters for the earliest O-type supergints are given in Table \ref{tab:obstars}.

\vspace{2mm}

Twenty (20) stars were duplicated between the O- and B-type supergiant lists.  Each was classified by \citep{1978AAS...31..243R} as an early B-type without luminosity class then re-classified by a subsequent survey as a later O-type with luminosity class.  In \citet{2023AJ....166..214M} we gave preference to the spectral types with luminosity class over the classifications of \citet{1978AAS...31..243R}. The following stars have been removed from the B-type supergiant list but remain in the O-type supergiants list:
SK -67    4, SK -70   13, SK -69   29, SK -66   47,  SK -70   57, SK -70   66, SK -70   70, HD 269244, SV* HV  2543, SK -67  150, HD 271363, HD 271366, SK -67  250, SK -68 166, SK -70  117, SK -68 180, SK -67  118, HD 269215, SK -67  119, and BI 137.

\vspace{2mm}

The VFTS survey of the 30 Doradus region focused primarily on luminous early type stars but the initial sample for that survey was magnitude limited and included a number of late type stars (see Table 3 in \citet{2011AA...530A.108E}).  Combining information from that survey with \citet{1981ApJ...250..116M}, \citet{2018MNRAS.478.3138D}, and \citet{2019AA...624A.128B} six (6) stars were added to the RSG sample and one star was added to the A-type supergiants (Table \ref{tab:newlate}).

\vspace{2mm}

The star SK -70   13 / LB 3400 was listed twice in the B-supergiant list.  They appear to be the same star so the entry for LB 3400 was removed.

\vspace{2mm}

The star CPD -69 448 is misclassified in the Simbad database and \citet{2023AJ....166..214M}.  Simbad cites \citet{2015AA...574A..13E} as the source of the spectral type but CPD -69 448 = VFTS 271 is not typed in that publication.  \citet{1985AA...153..235M} classifies it as A Ib and \citet{2011AA...530A.108E} assigns it A7 II.  CPD -69 448 was moved from the B-type star list into the A-supergiant list.  The re-calculated parameters for it are given in Table \ref{tab:newlate}.

\startlongtable
\begin{deluxetable*}{cclrrrclcrcc}
\tabletypesize{\tiny}
\tablewidth{0pt} 
\tablenum{A1}
\tablecaption{Additional OB-Type Supergiants\label{tab:obstars}}
\tablehead{
&&&&&&\colhead{Phot}&&\colhead{SpType}\\
\colhead{RAJ2000} &\colhead{DecJ2000} & \colhead{Name} & 
\colhead{V} &\colhead{(B-V)}&\colhead{(U-B)}&
\colhead{Source\tablenotemark{a}}&\colhead{SpType}&
\colhead{Source\tablenotemark{a}}&\colhead{A$_V$\tablenotemark{b}}&
\colhead{Log(T$_{eff}$)}&\colhead{M$_{Bol}$}\\
}
\startdata
72.36675&-67.71119&SK -67 3&13.18&-0.23&-1.11&Z04&O7.5nfp&M17&0.18&4.54&-8.98\\
72.79409&-69.55573&[MNM2015] LMC156-2&13.65&-0.13&-1.15&Z04&O6.5 Ifp&M17&0.57&4.56&-8.81\\
74.14046&-66.47367&PGMW 1180&12.99&-0.17&-1.10&W99&O8.5II(f)&W99&0.35&4.51&-8.97\\
74.14213&-66.47381&PGMW 1191&14.90&-0.19&-1.02&W99&O9.5V&W99&0.25&4.50&-6.89\\
74.14313&-66.47439&Sch95 8&14.87&-0.28&-0.99&W99&O8.5V&W99&0.01&4.53&-6.86\\
74.14313&-66.47503&PGMW 1197&15.04&-0.24&-0.91&W99&B0V&W99&0.03&4.49&-6.46\\
74.14354&-66.47436&PGMW 1199&13.57&-0.21&-0.99&W99&O7.5III&W99&0.25&4.54&-8.66\\
74.14371&-66.47392&Sch95 6&14.77&-0.24&-0.96&W99&O9.5V&W99&0.09&4.50&-6.86\\
74.14417&-66.47378&Sch95 7&14.91&-0.27&-1.03&W99&O9V&W99&0.03&4.52&-6.75\\
74.14467&-66.47381&Sch95 1&12.56&-0.18&-1.01&W99&O9Ib&W99&0.31&4.50&-9.24\\
74.14475&-66.47408&Sch95 5&14.74&-0.23&-1.00&W99&O9.5V&W99&0.12&4.50&-6.92\\
74.14517&-66.47464&PGMW 1212&14.66&-0.26&-0.95&W99&O8V&W99&0.08&4.54&-7.22\\
74.14525&-66.47314&PGMW 1213&13.82&-0.26&-0.99&W99&O8V((f))&W99&0.08&4.54&-8.06\\
74.31329&-69.33900&SK -69 50&13.37&-0.16&-1.04&Z04&O7(n)(f)(p)&W10a&0.40&4.56&-8.93\\
74.68703&-66.20405&SK -66 44&13.12&-0.30&-0.90&Z04&O6.5(n)fp&W10a&0.04&4.58&-8.97\\
79.70625&-69.25167&2dFL51-106&12.84&-0.14&-&U00&O7(n)fp&W10a&0.46&4.54&-9.44\\
80.22792&-65.45500&SK -65 47&12.47&-0.20&-0.87&Z04&O4I(n)f+p&W10a&0.42&4.61&-10.19\\
82.25712&-68.53436&SK -68 96&13.26&-0.15&-1.08&Z04&O7Ianf&W10a&0.43&4.54&-8.98\\
82.78383&-68.61508&SK -68 112&12.82&-0.18&-&Z04&O7.5(n)(f)(p)&W10a&0.33&4.53&-9.27\\
82.96708&-67.57250&HD 269702&11.99&-0.30&-&H00&O8I(f)p&W10a&0.00&4.52&-9.67\\
83.38057&-67.48273&[MNM2015] LMCe136-2&14.69&-0.16&-1.15&Z04&O6nfp&N18&0.47&4.59&-7.90\\
83.99421&-69.19775&H-M 6&14.14&0.04&-0.92&Z04&O8V&W99&1.01&4.54&-8.67\\
83.99867&-69.19704&H-M 8&14.30&0.01&-0.88&W99&O8III&W99&0.91&4.53&-8.34\\
83.99896&-69.19793&H-M 11A&15.15&0.09&-0.74&W99&O7.5V&W99&1.17&4.55&-7.91\\
83.99904&-69.19733&H-M 7&14.01&0.01&-1.03&W99&O4III&W99&1.05&4.63&-9.42\\
84.01892&-69.19725&H-M 9A&13.42&-&-&W99&O4If+&W99&0.00&4.61&-10.00\\
84.02667&-69.19667&SK -69 212&12.42&-0.02&-&Z04&O5n(f)p&W10a&0.95&4.59&-10.64\\
84.36637&-69.39789&UCAC4 104-015116&13.38&-0.16&-1.08&Z04&O5nfp&M17&0.49&4.61&-9.38\\
84.44258&-69.15283&Testor 1998 1B&15.10&0.20&-1.00&W99&O7V&W99&1.52&4.57&-8.39\\
84.44267&-69.15317&Testor 1998 1C&15.40&0.03&-0.58&W99&B1-2&W99&0.64&4.30&-5.63\\
84.44392&-69.15230&Testor 1998 2B&15.19&0.18&-1.23&W99&O8III&W99&1.44&4.53&-7.98\\
84.44417&-69.15222&[TLD88] 2A&14.13&0.13&-&W99&O4(n)(f)p&W99&1.43&4.61&-9.54\\
84.63750&-69.08700&Cl* NGC 2070 SMB 131&15.26&-0.06&-0.80&VFTS,P93a&O9.5(n)&W14&0.65&4.50&-6.93\\
84.66370&-69.11090&Cl* NGC 2070 SMB 365&16.38&0.54&-0.37&VFTS,Z04&O5:Vn&H12&2.67&4.61&-8.56\\
84.67010&-69.10130&Cl* NGC 2070 MH 95&15.38&0.39&-&VFTS&O9.5III/V&H12&2.05&4.50&-8.21\\
84.67120&-69.10050&Cl* NGC 2070 MH 129&14.68&0.17&-&VFTS&O6.5III/V&H12&1.46&4.58&-8.80\\
84.67180&-69.09780&[P93] 841&15.53&0.17&-1.10&P93a&O4-6(n) (f)p&W10a&1.52&4.61&-8.61\\
84.67520&-69.10470&Cl* NGC 2070 MH 314&15.51&0.15&-&VFTS&O3-4&H12&1.49&4.63&-8.50\\
84.67680&-69.10180&Cl* NGC 2070 MH 623&15.20&0.30&-&VFTS&O8.5III/V&H12&1.81&4.52&-8.19\\
84.67780&-69.10310&Cl* NGC 2070 MH 591&15.26&0.14&-&VFTS&O8III/V&H12&1.36&4.54&-7.63\\
84.67860&-69.08750&[P93] 994&15.54&0.01&-0.86&P93a&B0:V&SIM&0.83&4.49&-6.61\\
84.68030&-69.10459&Cl* NGC 2070 MH 749&13.80&0.04&-&VFTS&O3 III(f*) or O4-5V&W14&1.02&4.67&-10.04\\
84.68160&-69.09150&Cl* NGC 2070 SMB 240&16.13&-0.02&-1.04&VFTS,Z04&B0-0.5V&E15b&0.74&4.49&-5.84\\
84.68400&-69.09910&Cl* NGC 2070 MH 859&13.76&0.12&-0.62&P93a&O3III(f+)&M05a&1.40&4.67&-10.35\\
84.68430&-69.09640&Cl* NGC 2070 MH 878&13.36&1.70&-&VFTS&O8II&SIM&1.27&4.53&-9.57\\
84.69080&-69.09470&[CHH92] 7012&14.73&0.12&-1.04&P93a&O7&SIM&1.27&4.57&-8.43\\
84.69520&-69.10740&Cl* NGC 2070 SMB 408&16.73&0.17&-2.47&VFTS,Z04&B0vn&E15b&1.34&4.49&-5.84\\
84.71570&-69.10040&Cl* NGC 2070 SMB 229&15.96&0.04&-0.67&VFTS,Z04&B0-0.5V(n)&E15b&0.86&4.45&-5.92\\
\enddata
\tablenotetext{a}{E15a = \citet{2015AA...574A..13E}, H00 = \citet{2000AA...355L..27H}, H12 = \citet{2012AA...546A..73H}, M05a = \citet{2005ApJ...627..477M}, M17 = \citet{2017ApJ...837..122M}, N18 =\citet{2018ApJ...863..181N}, P93a = \citet{1993AJ....106.1471P}, SIM = classification given in SIMBAD which \citet{2000AAS..143....9W} identify as either from \citet{1975mcts.book.....H} or \citet{1978csst.book.....J}, VFTS = \citet{2011AA...530A.108E}, U00 = \citet{2000AcA....50..307U}, W10a = \citet{2010AJ....139.1283W}, W14 = \citet{2014AA...564A..40W}, W99 = \citet{1999AJ....118.1684W}, Z04 = \citet{2004AJ....128.1606Z}.}
\tablenotetext{b}{Extinction estimated from spectral type.}
\end{deluxetable*}

\begin{deluxetable*}{cclrrrrrlrcr}
\tabletypesize{\tiny}
\tablewidth{0pt} 
\tablenum{A3}
\tablecaption{Additional Late-Type Supergiants in 30 Doradus\label{tab:newlate}}
\tablehead{
\colhead{RAJ2000} &\colhead{DecJ2000} & \colhead{Name} &\colhead{J\tablenotemark{a}}&\colhead{H\tablenotemark{a}}&\colhead{K$_s$\tablenotemark{a}}&\colhead{V\tablenotemark{b}} &\colhead{(B-V)\tablenotemark{b}}&\colhead{SpType}&\colhead{A$_V$\tablenotemark{f}}&\colhead{Log(T$_{eff}$)\tablenotemark{d}}&\colhead{M$_{Bol}$\tablenotemark{f}}\\
}
\startdata 
\multicolumn{12}{c}{Red Supergiants}\\
\hline
84.566667&-69.16983056&V* Z Dor&9.308&8.336&7.867&14.22&2.28&M4 Ia-Ib\tablenotemark{3}&0.71&3.55&-7.60\\
84.569500&-69.07058056&2MASS J05381667-6904140&9.786&8.906&8.552&13.99&2.40&M\tablenotemark{4}&0.50&3.57&-7.08\\
84.611208&-69.14797222&WHO S 452&9.782&8.754&8.34&14.02&2.46&K5 Ia-Ib\tablenotemark{3}&0.66&3.57&-7.25\\
84.701917&-69.09241389&W61 7-8&9.173&8.297&7.86&13.62\tablenotemark{2}&2.31\tablenotemark{2}&M3.5 Ia\tablenotemark{3}&0.63&3.57\tablenotemark{e}&-7.75\\
84.914208&-69.19779167&W61 8-4&9.864&8.965&8.524&14.52&2.20&Early M\tablenotemark{5}&2.50\tablenotemark{6}&3.59&-7.58\tablenotemark{6}\\
84.924083&-69.19194722&W61 8-7&10.28&9.352&8.935&14.64&2.16&M\tablenotemark{4}&0.63&3.60&-7.75\\
\hline
\multicolumn{12}{c}{A-type Supergiants}\\
\hline
84.564040&-69.0675800&CPD-69 448&10.865&10.279&10.06&12.24&0.21&A7 II\tablenotemark{5}&0.27\tablenotemark{e}&3.92\tablenotemark{e}&-6.51\tablenotemark{e}\\
84.908000&-69.19613611&VFTS 820&10.865&10.279&10.06&12.73\tablenotemark{1}&0.55\tablenotemark{1}&A0 Ia\tablenotemark{5}&1.70\tablenotemark{e}&3.98\tablenotemark{e}&-7.51\tablenotemark{e}\\
\enddata
\tablenotetext{a}{From 2MASS point source catalog \citep{2003yCat.2246....0C}.}
\tablenotetext{b}{From the VFTS catalog \citep{2011AA...530A.108E} unless otherwise noted.}
\tablenotetext{d}{Estimated by \citet{2019AA...624A.128B} unless otherwise noted.}
\tablenotetext{e}{Estimated using the method outlined in \citet{2023AJ....166..214M}}
\tablenotetext{f}{Estimated by \citet{2018MNRAS.478.3138D} unless otherwise noted.}
\tablenotetext{1}{\citet{2002ApJS..141...81M}}
\tablenotetext{2}{\citet{1993AJ....106..560P}}
\tablenotetext{3}{\citet{2015AA...578A...3G}}
\tablenotetext{4}{\cite{1981ApJ...250..116M}}
\tablenotetext{5}{\citet{2011AA...530A.108E}}
\tablenotetext{6}{\citet{2019AA...624A.128B}}
\end{deluxetable*}

\begin{deluxetable*}{lc|cc|cc|cc}
\tabletypesize{\scriptsize}
\tablewidth{0pt} 
\tablenum{A4}
\tablecaption{Adopted Values by Spectral Type\label{tab:cal}}
\tablehead{
\multicolumn{2}{c}{ }&\multicolumn{2}{c}{\em{Supergiants (I)}}&\multicolumn{2}{c}{\em{Giants (III)}}&\multicolumn{2}{c}{\em{Dwarfs (V)}}\\
\colhead{Sp. Type}&\colhead{(B-V)$_0$}&
\colhead{Log(T$_{eff}$)}&\colhead{B.C (mags)}&
\colhead{Log(T$_{eff}$)}&\colhead{B.C (mags)}&
\colhead{Log(T$_{eff}$)}&\colhead{B.C (mags)}
}
\startdata 
O2&-&-&-&4.695&-4.34&4.727&-4.58\\
O3&-0.33&4.626&-3.87&4.672&-4.20&4.663&-4.12\\
O4&-0.33&4.607&-3.74&4.628&-3.88&4.632&-3.91\\
\enddata
\end{deluxetable*}

\movetabledown=2in
\begin{rotatetable}
\begin{deluxetable*}{cclrrrcclcrcc}
\tabletypesize{\scriptsize}
\tablewidth{0pt} 
\tablenum{A2}
\tablecaption{Additional OBA-Type Known and Suspected Binaries\label{tab:obabin}}
\tablehead{
&&&&&&
\colhead{Phot}&\colhead{Binary}&&\colhead{SpType}\\
\colhead{RAJ2000} & \colhead{DecJ2000}& \colhead{Name} & \colhead{V} &\colhead{(B-V)}&\colhead{(U-B)}&
\colhead{Source\tablenotemark{a}}&\colhead{Class\tablenotemark{c}}&
\colhead{SpType}&\colhead{Source\tablenotemark{a}}&\colhead{A$_V$\tablenotemark{b}}&\colhead{Log(T$_{eff}$)}&\colhead{M$_{Bol}$}\\
}
\startdata 
74.14054&-66.47328&PGMW 1181&14.19&-0.24&-0.99&W99&SB&O9Vn(sb?)&W99&0.124&4.52&-7.56\\
74.142047&-66.47383&OGLE LMC-ECL-27754&15.08&-0.28&-0.83&W99&SB&B1V&W99&0.213&4.36&-6.60\\
74.14354&-66.47411&PGMW 1198&14.88&-0.27&-0.92&W99&Blend&B0.5-B1V&W99&0.213&4.39&-6.17\\
84.44200&-69.15246&VFTS 145&14.30&0.19&-&VFTS&SB&O8fp&W14&1.49&4.52&-9.36\\
84.44338&-69.15338&IRSF J05374638-6909120&15.30&0.12&-0.11&W99&SB&O9.5V&W99&1.209&4.50&-7.45\\
84.44463&-69.15313&IRSF J05374670-6909112&15.30&0.06&-1.06&W99&Blend&O7V&W99&1.085&4.57&-7.76\\
84.66250&-69.11640&Cl* NGC 2070 SMB 278&15.86&0.44&0.63&VFTS,M02&SB&O7.5V+O7.5V&W14&2.26&4.55&-8.15\\
84.66540&-69.10240&[P93] 729&15.26&0.10&-1.07&VFTS,P93a&Blend&O6-7dbl&H12&1.24&4.58&-8.00\\
84.67120&-69.08700&Cl* NGC 2070 SMB 81&14.66&-0.06&-&VFTS&Blend&O3.5V((f))z+OB&W14&0.84&4.64&-8.73\\
84.67298&-69.11383&Cl* NGC 2070 MH 203&13.67&0.61&-&VFTS&Blend&O3 V + mid/late O&W14&2.91&4.66&-10.82\\
84.67390&-69.10530&Cl* NGC 2070 SMB 154&16.55&0.17&-2.14&VFTS,Z04&SB&O8.5 V&W14&1.72&4.53&-7.16\\
84.67446&-69.10346&Cl* NGC 2070 MH 290&14.34&0.28&-&VFTS&Blend&O3 III(f*) + mid/late O&W14&1.77&4.63&-9.33\\
84.67528&-69.10394&Cl* NGC 2070 SMB 25&13.31&0.21&-&VFTS&Blend&O4 If+  &W14&1.67&4.61&-11.04\\
84.67917&-69.07042&MH 12&13.96&-0.03&-2.32&VFTS,M02&SB&ON9Ia:+O7.5:I:(f):&W14&0.78&4.50&-8.79\\
84.68480&-69.09850&Cl* NGC 2070 MH 887&13.55&1.70&-0.79&P93a&SB&OC2.5If*+O4V&B22&1.55&4.63&-10.43\\
84.68560&-69.09480&Cl* NGC 2070 MH 917&17.10&0.18&-&VFTS&SB&O9.5II-III(n)&W14&1.40&4.49&-5.92\\
\enddata
\tablenotetext{a}{B22= \citet{2022MNRAS.510.6133B}, H12 = \citet{2012AA...546A..73H}, M02 = \citet{2002ApJS..141...81M}, P93a = \citet{1993AJ....106.1471P}, VFTS = \citet{2011AA...530A.108E}, W14 = \citet{2014AA...564A..40W}, W99 = \citet{1999AJ....118.1684W}, Z04 = \citet{2004AJ....128.1606Z}.}
\tablenotetext{b}{Estimated from spectral type}
\tablenotetext{c}{Multiplicity types: Blend = unresolved blend of more than one star, SB = Spectroscopic Binary}
\end{deluxetable*}
\end{rotatetable}

\section{Maps of Luminous Stars in the LMC} \label{lmcmaps}
\begin{figure}
\epsscale{0.44}
\figurenum{B1}
\plotone{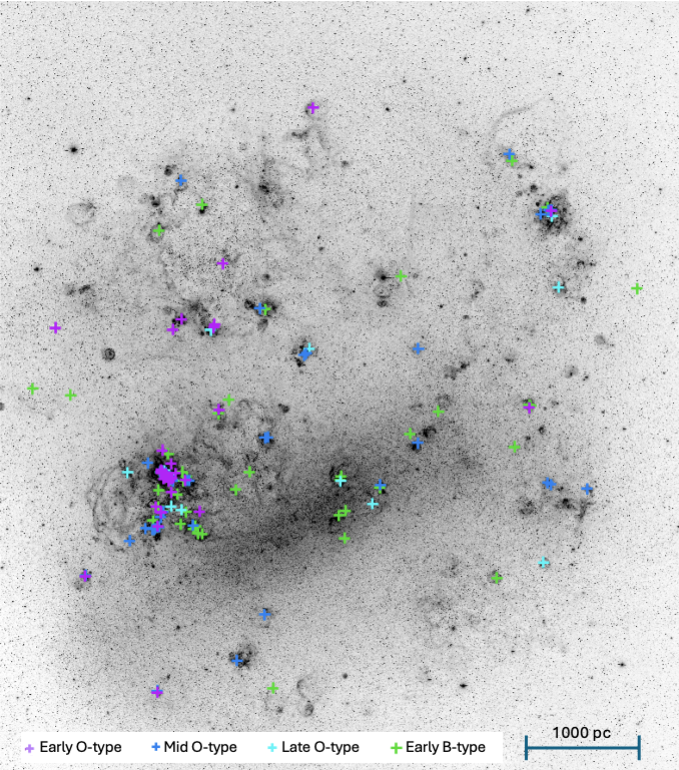}
\caption{An H-alpha emission image of the LMC from MCELS \citep{1999IAUS..190...28S} for the high luminosity reference samples with $M_{Bol} < -9.7$ (Table \ref{tab:refsamp}): Early O-types (purple), Mid O-types (blue), Late O-types (cyan), and Early B-types (green).}
\label{fig:LMCMap_hilum}
\end{figure}

\begin{figure}
\epsscale{0.43}
\figurenum{B2}
\plotone{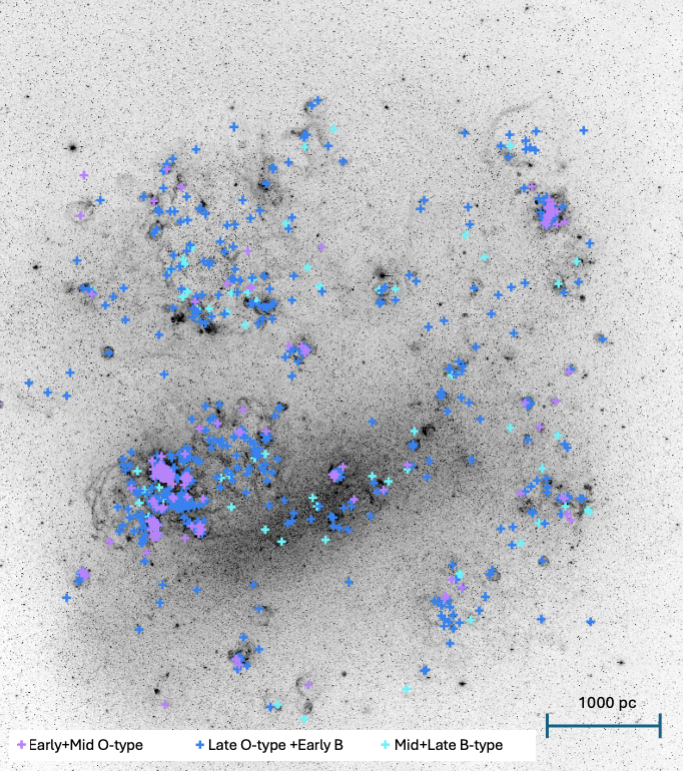}
\caption{An H-alpha emission image of the LMC from MCELS \citep{1999IAUS..190...28S} including the reference samples of the earliest types for $-8 > M_{Bol} <\geq -9.7$ (Table \ref{tab:refsamp}): Early + Mid O-types (purple), Late O-types + Early B-types (blue), and Mid + Late B-types (cyan).}
\label{fig:LMCMap_midlum_early}
\end{figure}

\begin{figure}
\epsscale{0.48}
\figurenum{B3}
\plotone{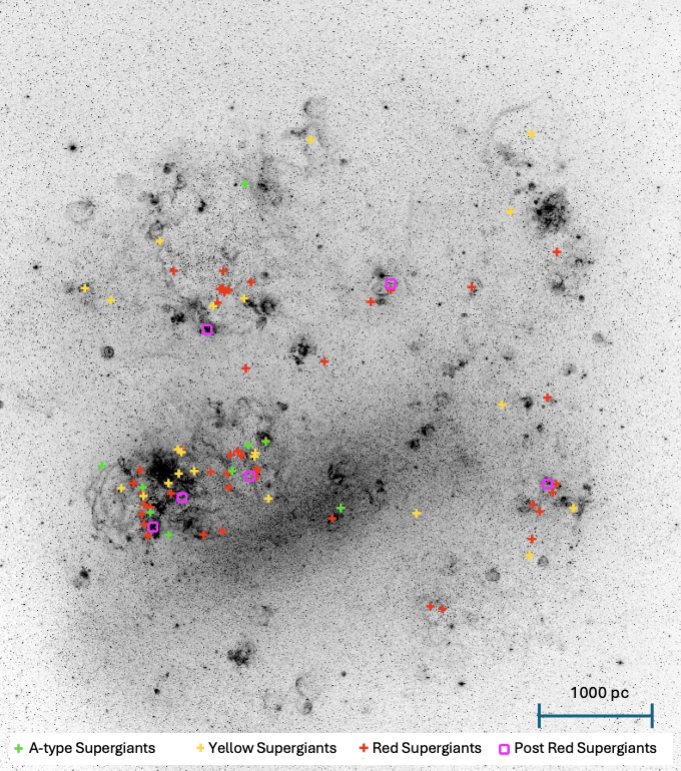}
\caption{An H-alpha emission image of the LMC from MCELS \citep{1999IAUS..190...28S} including the reference samples of the late types for $-8 > M_{Bol} \geq -9.7$ (Table \ref{tab:refsamp}): A-type supergiants (green), Yellow supergiants (gold), Red supergiants (red) and post-Red supergiants (magenta squares).}
\label{fig:LMCMap_midlum_late}
\end{figure}

\begin{figure}
\epsscale{0.49}
\figurenum{B4}
\plotone{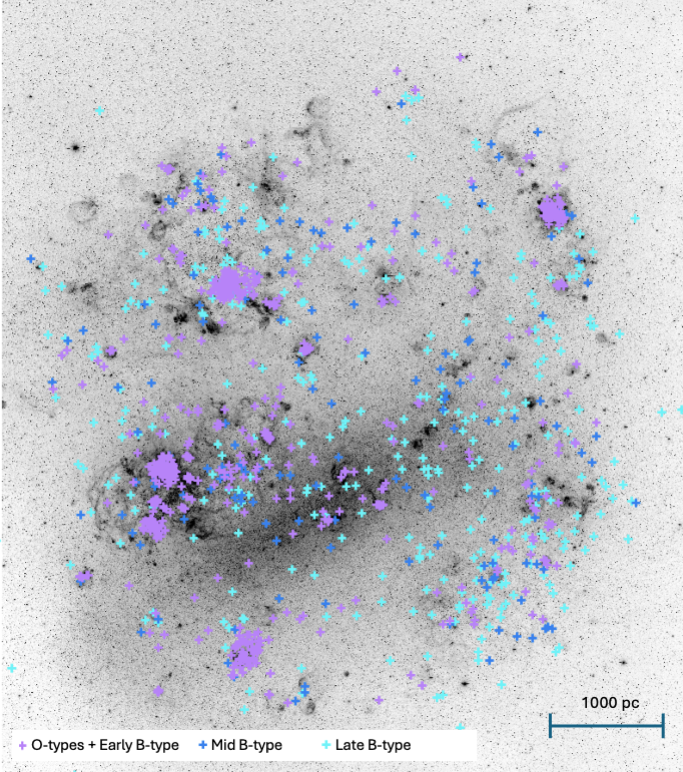}
\caption{An H-alpha emission image of the LMC from MCELS \citep{1999IAUS..190...28S} including the reference samples for $-5.0 > M_{Bol} \geq -7.0$ (Table \ref{tab:refsamp}): O-types + Early B-types (purple), Mid B-types (blue), and Late B-types (cyan).}
\label{fig:LMCMap_lowlum_early}
\end{figure}

\begin{figure}
\epsscale{0.49}
\figurenum{B5}
\plotone{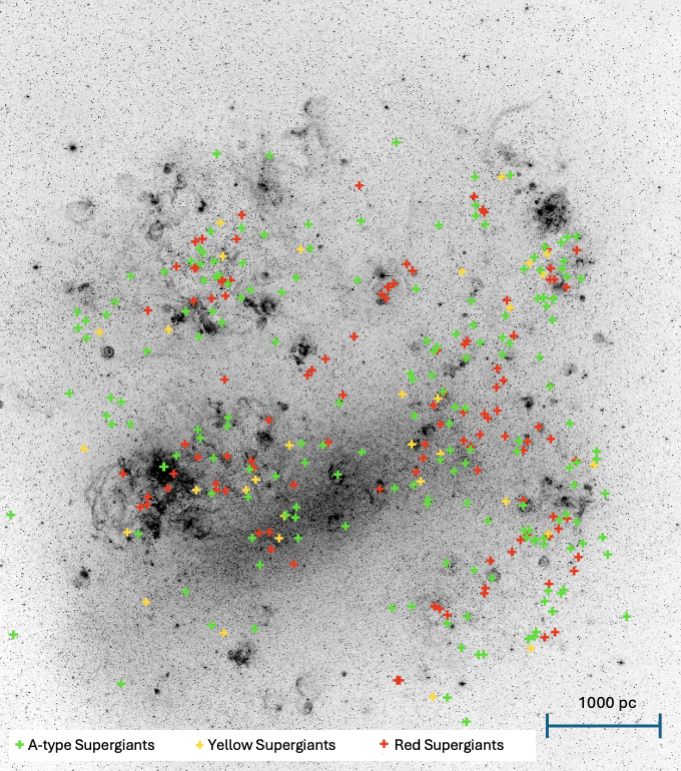}
\caption{An H-alpha emission image of the LMC from MCELS \citep{1999IAUS..190...28S} including the reference samples of the late types for $-5 > M_{Bol} \geq -7.0$ (Table \ref{tab:refsamp}): A-type supergiants (green), Yellow supergiants (gold), and Red supergiants (red).}
\label{fig:LMCMap_lowlum_late}
\end{figure}

In Figures  \ref{fig:LMCMap_hilum} -- \ref{fig:LMCMap_lowlum_late}, the reference samples (Table \ref{tab:refsamp}) are plotted on  maps of H-alpha emission from the UM/CTIO Magellanic Cloud Emission-line Survey (MCELS) \citep{1999IAUS..190...28S}.

\section{Wolf-Rayet Stars Samples} \label{wrlisting}
The list of WR in the LMC is taken from \citet{2018ApJ...863..181N}. 
The WN and WNh are split into higher and lower luminosity, above and below the Humphreys-Davidson limit ($M_{Bol} = -9.7$).  $M_{Bol}$ values for most of the WR are given in \citet{2014AA...565A..27H}.   Five WN have no $M_{Bol}$ determined by \citet{2014AA...565A..27H}.  [M2002]LMC 57799, [M2002]LMC 169271, and [MNM2014]LMC143-1 are all WN3 stars discussed by \citet{2014ApJ...788...83M}. Comparing their $M_V$ to other WN3 with known $M_{Bol}$, [M2002]LMC 57799 and [M2002]LMC 169271 are classified lower-luminosity and [MNM2014]LMC143-1 is classified high-luminosity.  Brey 3a and LHA 120-S 131 are classified lower-luminosity based on analyses by \citep{1991A&A...244L...9M} and \citep{2015ApJ...807...81M} respectively.  See Table \ref{tab:wr_wn}.

\vspace{2mm}

Fifteen (15) of the 23 WC stars do not have $M_{Bol}$ determined by previous work.  The rest range between $-8.9 \geq M_{Bol} \geq -9.9$.  The absolute visual magnitude $M_V$ was estimated for for 18 of the WC using the observed V magnitude and visual extinction ($A_v$) determined by \citet{2023MNRAS.521..585C}.  Eight out of eleven brightest stars with $M_V < -5.9$ have composite spectra indicating unresolved companions may be contributing to their brightness.  The 13 WC stars without composite spectra range from $-4.67 \geq M_V \geq -6.04$.  WC in the LMC all have the same sub type, a similar metalicity \citep{1990ApJ...348..471S} and cover a relatively narrow range of brightness when accounting for unresolved companions. So they are kept together as a single sample.

\vspace{2mm}

The WO, WN3/O3 and Of/WN samples are also {\em not} split by luminosity. The WO and WC stars share a suspected evolutionary link \citep{2022ApJ...931..157A} but all 3 of the WO stars are lower luminosity (Table \ref{tab:wr_wc}).  WN3/O3 are a sub-class of WN3 and WN4 with some hydrogen in their spectra \citep{2017ApJ...841...20N}. Not to be confused with the Of/WN "slash" stars, the WN3/O3 are all lower luminosity with $M_{Bol} \geq -9.8$ (Table \ref{tab:wr_wn}).  Slash-stars (Of/WN) are a classification created by \citet{2011MNRAS.416.1311C} to describe high-luminosity ($M_{Bol} \leq -9.7$) stars that show a combination of Of and WN features in their spectra (Table \ref{tab:wr_slash}). 

\begin{figure}[b]
\epsscale{0.55}
\figurenum{C1}
\plotone{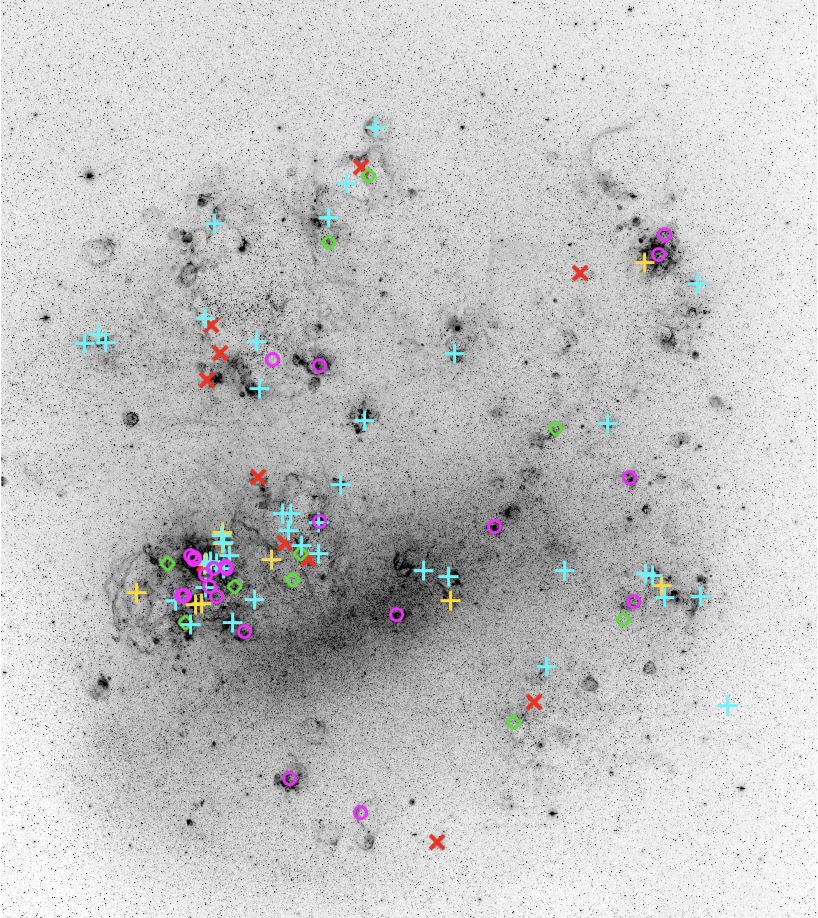}
\caption{An H-alpha emission image of the LMC from MCELS \citep{1999IAUS..190...28S} with earlier type WN (WN2-WN4, cyan crosses), later type WN (WN5-WN11, yellow crosses), WNh (red X's), and WN3/O3 (green diamonds) in the luminosity range from $-9.7 \leq M_{Bol} < -8$. WC (magenta circles) of all luminosity types are also plotted.}
\label{fig:wr_low_lum_Map}
\end{figure}

\begin{figure}
\epsscale{0.49}
\figurenum{C2}
\plotone{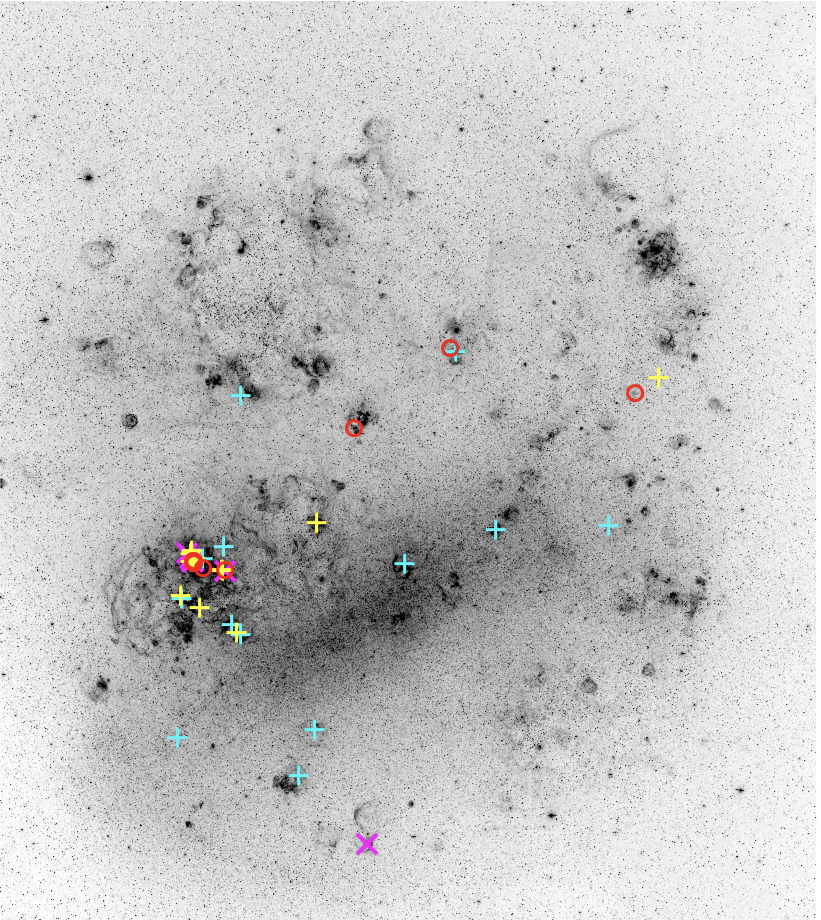}
\caption{An H-alpha emission image of the LMC from MCELS \citep{1999IAUS..190...28S} with earlier type WN (WN3-WN4, cyan crosses), later type WN (WN4.5-WN7, yellow crosses), WNh (magenta X's) and Of/WN slash stars (red circles) in the highest luminosity range for stars brighter than $M_{Bol} < -9.7$. Note that all but one of the WNh are found in and around the 30 Doradus region.}
\label{fig:wr_hi_lum_Map}
\end{figure}

Figures \ref{fig:wr_low_lum_Map} and \ref{fig:wr_hi_lum_Map} show the distributions of the  WR samples plotted on a map of the LMC.

\startlongtable
\begin{deluxetable*}{cclcccc}
\tabletypesize{\tiny}
\tablewidth{0pt} 
\tablenum{C1}
\tablecaption{Wolf-Rayet WN, WNh, and WN3/O3 in the LMC\label{tab:wr_wn}}
\tablehead{
\colhead{RAJ2000} &\colhead{DecJ2000} & \colhead{Name} &\colhead{SpType}&\colhead{Log(T$_{eff}$)}&\colhead{M$_{Bol}$}&\colhead{Source\tablenotemark{a}}\\
}
\startdata 
\hline\\
\multicolumn{7}{c}{WN ($M_{Bol} < -9.7$)}\\
\hline\\
73.8806250&-67.5007500&HD 32109&WN3pec&5.20&-9.80&1\\
74.7363333&-68.8030556&HD 268856&WN4+OB&4.83&-9.85&1\\
77.4184167&-68.8902222&HD 34169&WN3+OB&4.70&-10.55&1\\
78.4323750&-67.3748333&HD 34632&WN4+OB&4.83&-10.95&1\\
79.5800417&-69.1946389&HD 269333&BI+WN4&4.85&-13.45&1\\
81.6888333&-68.8313333&HD 269546&B3I+WN5&4.85&-15.20&1\\
81.9070000&-70.6015000&Brey 37&WN3+OB&4.83&-9.83&1\\
82.3883750&-70.9930278&Brey 40a&WN3+O7.5&4.85&-11.05&1\\
83.2940417&-67.7119722&Brey 48&WN3+OB&4.85&-11.33&1\\
83.6503333&-69.7601389&Sk-69 194&B0I+WN3&-&-10.50&2\\
83.7474167&-69.7350833&Brey 53&WN3+O&4.85&-10.33&1\\
83.8688333&-69.6691389&[MNM2014]LMC143-1&WN3+O8-9III&-&high&\\
83.9345000&-68.9935556&Brey 60&WN3+abs&4.80&-10.15&1\\
83.9952917&-69.1966111&Brey 65&WN7&4.65&-12.18&1\\
83.9992500&-69.1895278&Brey 57&WN7&4.62&-10.63&1\\
83.9995417&-69.1973889&BAT99 80&WN5 g&4.65&-11.20&1\\
84.4543333&-69.0856389&HD 269891&B1I+WN3&4.65&-12.58&1\\
84.6400833&-69.0806944&HD 269919&WN7&4.62&-10.20&1\\
84.6517500&-69.1159444&Brey 81&WN7&4.67&-11.08&1\\
84.6631250&-69.1058889&Brey 79&WN6&4.65&-11.95&1\\
84.6734167&-69.0875278&RMC140b&WN5(h)+O&4.67&-10.35&1\\
84.6980000&-69.0070278&HD 269926&WN4.5&4.80&-11.20&1\\
84.7224167&-69.0335833&HD 38282 &WN5/6+WN6/7 i&4.67&-11.85&1\\
84.7377917&-69.1015833&HD 269928&WN6+O3.5If*/WN7&4.67&-11.63&1\\
85.0314583&-69.4088611&HD 38472&WN3+O7&4.85&-11.30&1\\
85.0555417&-69.3795833&HD 38489&B[e]+WN?&4.41&-9.98&4\\
85.4523750&-70.5918889&Brey 97&WN3+O5V&4.90&-10.70&1\\
\hline\\
\multicolumn{7}{c}{WN ($ -9.7 \leq M_{Bol} < -8$)}\\
\hline\\
71.3843750&-70.2530278&Brey 1&WN3&4.95&-8.45&1\\
72.4010417&-69.3485556&Brey 2&WN2&5.15&-8.63&1\\
73.2391250&-66.6870833&Brey 3&WN3&4.90&-8.98&1\\
73.2656667&-69.3977222&[M2002]LMC 15666&WN3+O6V&4.95&-8.83&5\\
73.3753333&-69.2970556&Brey 3a&WNL/Of?&-&low&\\
73.6170833&-69.2141389&Brey 4&WN2&5.15&-8.83&1\\
73.7816667&-69.2088056&[M2002]LMC 23417&WN3&-&-8.80&6\\
74.4210000&-66.5451667&LHA 120-S 9&WN10&4.85&-9.10&1\\
74.9647917&-67.9487222&HD 268847&WN3&4.95&-9.13&1\\
75.7468333&-69.2339722&[M2002]LMC 57799&WN3+abs&-&low&\\
76.0513750&-70.0653889&Brey 14&WN4&4.83&-9.43&1\\
78.4761250&-69.5296389&Brey 18&WN9&4.51&-9.58&1\\
78.4833750&-67.4101667&BAT99 23&WN3&4.85&-9.08&1\\
78.5528750&-69.3239444&Brey 19&WN3&5.00&-9.05&1\\
79.1618333&-69.2780278&Brey 20&WN4&4.85&-9.25&1\\
80.1863750&-65.4723611&Brey 23&WN3+OB&4.85&-8.95&1\\
80.5183750&-67.9852222&Brey 25&WN3&4.88&-8.53&1\\
80.8250417&-65.9491667&Brey 27&WN3&4.85&-9.20&1\\
81.1007917&-68.5265556&Brey 29&WN3/WCE+OB&4.85&-9.48&1\\
81.2264167&-66.2364167&Brey 30&WN3&4.90&-9.33&1\\
81.6535833&-68.8503611&Brey 33&WN4&4.45&-9.25&1\\
81.6774167&-69.1159444&HD 269549&WN4&5.00&-9.20&1\\
82.0745833&-69.0433056&Brey 38&WN4&4.80&-8.80&1\\
82.3015417&-68.7600278&HD 269618&WN3&4.95&-9.18&1\\
82.3818333&-68.9080000&HD 269624&WN3&4.95&-8.70&1\\
82.5102500&-68.7551111&Brey 42&WN3&4.95&-8.45&1\\
82.8252083&-69.1459722&Brey 44a&WN9&4.58&-9.58&1\\
82.8931667&-67.2748056&Brey 45&WN3&4.90&-8.70&1\\
83.2952917&-69.4836111&Brey 49&WN3+OB&4.80&-9.65&1\\
83.6561250&-66.2438889&Brey 51&WN3&4.85&-8.73&1\\
83.8132500&-69.0953056&Brey 55&WN3&4.83&-9.58&1\\
83.9376250&-68.9790000&Brey 61&WN4+abs&4.85&-9.70&1\\
83.9610417&-68.8942222&Brey 63&WN5&4.78&-9.50&1\\
83.9684583&-68.9190833&Brey 63a&WN3+abs&4.90&-9.43&1\\
83.9751250&-67.0469167&Brey 59&WN4&4.85&-9.10&1\\
83.9765417&-68.9855278&LHA 120-S 125&WN9&4.54&-9.35&1\\
83.9965000&-69.1974167&Brey 65b&WN4&4.85&-9.45&1\\
84.1399167&-69.1548056&Brey 66&WN3&5.00&-9.03&1\\
84.2978333&-69.1272778&Brey 69&WN3&4.95&-8.53&1\\
84.3988333&-69.1445278&Brey 70a&WN3/WCE&5.05&-9.70&1\\
84.4187500&-69.1326944&Brey 71&WN6&4.70&-9.65&1\\
84.4484167&-69.3537778&[M2002]LMC 169271&WN3+O7V&-&low&\\
84.6008750&-69.4870556&LHA 120-S 131&WN11&-&low&\\
84.6154583&-69.4995833&HD 269908&WN3/4pec&5.15&-9.70&1\\
84.7420417&-69.4887500&Brey 91&WN9&4.51&-9.15&1\\
84.9007500&-69.6531111&Brey 93a&WN3&4.80&-8.83&1\\
85.2116667&-69.4421667&Brey 96&WN3&5.05&-8.80&1\\
86.1292917&-69.3376389&Sk-69 296&WN11&4.45&-9.40&1\\
86.2238333&-67.1767222&Brey 98&WN4&4.85&-9.38&1\\
86.3506667&-67.0991111&Brey 99&WN4&4.90&-9.15&1\\
86.6931250&-67.1661944&HD 270149&WN3&4.90&-8.98&1\\
\multicolumn{7}{c}{WNh ($M_{Bol} < -9.7$)}\\
\hline\\
80.5938750&-71.5994722&HD 36063&WN6h&4.67&-10.05&1\\
83.9257917&-69.2095833&Brey 56&WN5h&4.67&-10.10&1\\
84.6689625&-69.0991944&Brey 75&WN6h&4.67&-10.58&1\\
84.6733333&-69.0872222&RMC140a&WN6&4.65&-12.20&1\\
84.6763875&-69.1009167&BAT99 106&WN4.5h&4.75&-10.98&1\\
84.6766250&-69.1008056&BAT99 108&WN5h&4.75&-12.38&1\\
84.6767167&-69.1007778&BAT99 109&WN5h&4.75&-10.83&1\\
84.6787750&-69.1013333&BAT99 112&WN4.5h&4.75&-11.40&1\\
84.6844042&-69.1016111&Brey 84&WN4.5h&4.80&-12.83&1\\
84.7313750&-69.0740833& VFTS682&WN5h&4.72&-11.45&3\\
84.7972083&-69.0337778&HD 38344&WN5h&4.70&-10.78&1\\
\multicolumn{7}{c}{WNh ($ -9.7 \leq M_{Bol} < -8$)}\\
\hline\\
75.7871250&-66.6826389&Brey 13&WN7h&4.70&-9.70&1\\
76.2851250&-70.3791944&Brey 15&WN3h&4.85&-9.28&1\\
78.7386250&-71.6050833&Brey 19a&WN4ha&4.83&-9.08&1\\
80.4904167&-65.8167500&Brey 24&WN6h&4.67&-9.33&1\\
81.9278750&-69.1667778&Brey 36&WN7h&4.65&-9.35&1\\
82.4735000&-69.0180000&Brey 41&WN5h&4.81&-9.33&1\\
83.0312083&-68.4421111&Brey 47&WN7h&4.67&-9.30&1\\
83.7167917&-67.3580556&Brey 52&WN4h&4.80&-9.15&1\\
83.8741667&-67.1137222&Brey 54&WN3(h)&4.95&-9.65&1\\
84.0505417&-67.5827222&Brey 65a&WN5h&4.67&-8.90&1\\
84.4431250&-69.1526667&Brey 73&WN6h&4.65&-8.75&1\\
\multicolumn{7}{c}{WN3/O3 ($ -9.7 \leq M_{Bol} < -8$)}\\
\hline\\
74.2032917&-69.6113056&[MNM2015]LMCe055-1&WN4/O4&4.97&-7.63&8\\
76.1360000&-68.0165000&[MNM2014]LMC277-2&WN3/O3&5.00&-9.70&7\\
76.8055417&-70.5594167&[M2002]LMC 71747&WN3/O3&4.98&-9.18&7\\
80.3450833&-65.8802222&[MNM2015]LMCe169-1&WN3/O3&5.00&-8.28&7\\
81.2369583&-66.4456667&[MNM2015]LMCe159-1&WN3/O3&5.01&-9.53&7\\
82.1130000&-69.1100556&[MNM2014]LMC199-1&WN3/O3&5.00&-8.88&7\\
82.3257917&-69.3286667&[M2002]LMC 143741&WN3/O3&5.00&-9.43&7\\
83.7537500&-69.3556111&[MNM2014]LMC172-1&WN3/O3&5.02&-9.85&7\\
85.0148750&-69.6314167&[MNM2014]LMC174-1&WN3/O3&5.00&-8.88&7\\
85.3229167&-69.1156111&[MNM2015]LMCe078-3&WN3/O3&5.00&-8.60&7\\
\enddata
\tablenotetext{a}{Source of $log(T_{eff})$ and $M_{Bol}$: 1 =\citet{2014AA...565A..27H}, 2 =\citet{2000AJ....119.2214M}, 3 =\citet{2014AA...570A..38B}, 4 =\citet{1998AAS..130..527T}, 5 =\citet{2014MNRAS.442..929G}, 6 =\citet{2023MNRAS.521..585C}, 7 =\citet{2017ApJ...841...20N}, 8 =\citet{2024ApJ...977...82M}}
\end{deluxetable*}

\begin{deluxetable*}{cclccccccc}
\tabletypesize{\tiny}
\tablewidth{0pt} 
\tablenum{C2}
\tablecaption{Wolf-Rayet WC and WO in the LMC\label{tab:wr_wc}}
\tablehead{
\colhead{RAJ2000} &\colhead{DecJ2000} & \colhead{Name} &\colhead{SpType}&\colhead{$V_{mag}$}&\colhead{$A_V$\tablenotemark{b}}&\colhead{$M_V$}
&\colhead{Log(T$_{eff}$)}&\colhead{M$_{Bol}$}&\colhead{Source\tablenotemark{a}}
}
\startdata 
\hline\\
\multicolumn{10}{c}{WC}\\
\hline\\
74.0120000&-69.4559722&HD 32257&WC4&14.23&0.4&-4.67&4.94&-8.90&9\\
74.0458333&-66.2925000&HD 32125&WC4&14.33&0.5&-4.67&4.92&-8.90&9\\
74.1442917&-66.4740000&HD 32228&WC4+OB&10.83&0.1&-7.77&-&-&\\
74.3504167&-68.3992500&HD 32402&WC4&12.96&0.4&-5.94&4.86&-9.85&9\\
77.4740417&-68.8812500&Brey 16a&WC4&13.74&1.2&-5.96&-&-&\\
79.8180833&-69.6555556&Brey 22&WC6+O5-6&12.21&0.2&-6.49&-&-&\\
80.7919167&-71.3474722& HD 36156&WC4+abs&12.66&0.1&-5.94&-&-&\\
81.5165000&-67.4991944&HD 36402&WC4+abs&11.62&0.2&-7.08&5.01&-9.53&7\\
81.6260833&-68.8409722&HD 36521&WC4+O6III/V&12.32&0.1&-6.28&-&-&\\
82.5506667&-67.4356389&HD 37026&WC4&13.56&0.3&-5.24&4.92&-9.33&9\\
82.6612500&-71.0299444&HD 37248&WC4+abs&12.96&0.4&-5.94&-&-&\\
83.5801667&-69.7528611&HD 37680&WC4&13.12&0.5&-5.88&4.86&-9.90&9\\
83.9248333&-69.1980278&Brey 58a&WC4&16.52&-&-&-&-&\\
83.9312083&-69.1827778&Brey 62&WC4&13.96&1.5&-6.04&-&-&\\
84.2140833&-69.4324167&HD 38030&WC4(+OB)&12.99&0.5&-6.01&-&-&\\
84.2277500&-69.1939722&Brey 67&WC4(+OB)&12.18&1.1&-7.42&-&-&\\
84.3718333&-69.3465278&Brey 70&WC4&13.69&-&-&-&-&\\
84.4360000&-69.2404722&HD 269888&WC4&14.63&1.1&-4.97&4.93&-8.93&9\\
84.6733333&-69.0872222&RMC140&WC4&12.2&-&-&-&-&\\
84.6835958&-69.0987500&Brey 83&WC5&14.47&1.6&-5.63&-&-&\\
84.7657500&-69.0629167&Brey 90a&WC4&15.83&2.2&-4.87&-&-&\\
84.9837917&-69.4067500&HD 38448&WC4+abs&13&-&-&4.88&-&4\\
85.0544167&-69.4011667&Brey 95a&WC4+O&13.3&-&-&-&-&\\
\hline\\
\multicolumn{10}{c}{WO}\\
\hline\\
79.5430417&-69.2173611&[MNM2014]LMC195-1&WO2&15.15&0.40&-3.75&5.05&-8.73&1\\
79.5453333&-69.2198333&[L72]LH41-1042&WO4&14.00&0.40&-4.9&-&-8.33&4\\
84.8928750&-68.7358889&Brey 93&WO3&15.20&0.50&-3.8&5.04&-8.73&1\\
\enddata
\tablenotetext{a}{Source of $log(T_{eff})$ and $M_{Bol}$: 1 =\citet{2022ApJ...931..157A}, 2 =\citet{1998AAS..130..527T}, 3 =\citet{2017ApJ...841...20N}, 4 =\citet{2023MNRAS.521..585C}}
\tablenotetext{b}{Derived by \citet{2023MNRAS.521..585C}.}
\end{deluxetable*}

\begin{deluxetable*}{cclccc}
\tabletypesize{\scriptsize}
\tablewidth{0pt} 
\tablenum{C3}
\tablecaption{Wolf-Rayet Slash Stars in the LMC\label{tab:wr_slash}}
\tablehead{
\colhead{RAJ2000} &\colhead{DecJ2000} & \colhead{Name} &\colhead{SpType}&\colhead{Log(T$_{eff}$)\tablenotemark{a}}&\colhead{M$_{Bol}$\tablenotemark{a}}
}
\startdata 
74.3643333&-67.6508056&Brey 10a&O2If*/WN5&4.70&-9.70\\
78.5732083&-67.3430833&[MNM2015]LMCe132-1&O3.5If*/WN5&-&\\
80.7490833&-68.0296111&HD 269445&Ofpe/WN9&4.45&-11.45\\
83.9258333&-69.1982222&Brey 58&O3.5If*/WN7&4.65&-10.20\\
84.4639167&-69.1629722&Brey 74a&O3If*&4.65&-9.95\\
84.6618333&-69.1137500&BAT99 97&O3.5If*/WN7&4.65&-10.95\\
84.6676167&-69.0999444&Brey 78&O2.5If*/WN6&4.65&-9.95\\
84.6744792&-69.1039722&Brey 76&O2If*/WN5&4.80&-11.48\\
84.6754833&-69.0986667&Brey 77&O2If*&4.70&-11.20\\
84.6767000&-69.1041667&4 R136-015&O2If*/WN5&-&-11.93\\
84.6767917&-69.1007500&BAT99 110&O2If*/O3If*/WN6&4.70&-10.75\\
84.6795833&-69.0963333&BAT99 113&O2If*/WN5&4.70&-10.43\\
84.6800458&-69.1040000&BAT99 114&O2If*/WN5&4.80&-11.30\\
\enddata
\tablenotetext{a}{$log(T_{eff})$ and $M_{Bol}$ from \citet{2014AA...565A..27H}}
\end{deluxetable*}

\vspace{10mm}
\bibliography{LMC_HRD_2}{}
\bibliographystyle{aasjournal}



\end{document}